\documentclass[twocolumn]{aastex63}

\newcommand{\dia}[0]{2.4\,m} 
\newcommand{\len}[0]{2.6\,m} 

\usepackage{multirow}
\usepackage{amsmath}

\usepackage{lineno}

\shorttitle{The Simons Observatory (SO) Large Aperture Telescope Receiver (LATR)}
\shortauthors{Zhu et al.}
\graphicspath{{./}{figures/}}

\begin{document}

\title{The Simons Observatory Large Aperture Telescope Receiver}

\correspondingauthor{Ningfeng Zhu}
\email{nin@sas.upenn.edu}

\author{Ningfeng Zhu}
\affiliation{Department of Physics and Astronomy, University of Pennsylvania, 209 South 33rd Street, Philadelphia, PA 19104, USA}

\author{Tanay Bhandarkar}
\affiliation{Department of Physics and Astronomy, University of Pennsylvania, 209 South 33rd Street, Philadelphia, PA 19104, USA}

\author{Gabriele Coppi}
\affiliation{Department of Physics and Astronomy, University of Pennsylvania, 209 South 33rd Street, Philadelphia, PA 19104, USA}
\affiliation{Department of Physics, University of Milano-Bicocca, Milano (MI), Italy}

\author{Anna M. Kofman}
\affiliation{Department of Physics and Astronomy, University of Pennsylvania, 209 South 33rd Street, Philadelphia, PA 19104, USA}

\author{John L. Orlowski-Scherer}
\affiliation{Department of Physics and Astronomy, University of Pennsylvania, 209 South 33rd Street, Philadelphia, PA 19104, USA}

\author{Zhilei Xu}
\affiliation{Department of Physics and Astronomy, University of Pennsylvania, 209 South 33rd Street, Philadelphia, PA 19104, USA}
\affiliation{MIT Kavli Institute, Massachusetts Institute of Technology, Cambridge, MA, USA}

\author{Shunsuke Adachi}
\affiliation{Department of Physics, Faculty of Science, Kyoto University, Kitashirakawa Oiwake-cho, Sakyo-ku, Kyoto 606-8502, Japan}

\author{Peter Ade}
\affiliation{School of Physics and Astronomy, Cardiff University, The Parade, Cardiff, CF24 3AA, UK}

\author{Simone Aiola}
\affiliation{Center for Computational Astrophysics, Flatiron Institute, 162 5th Avenue New York, NY 10010, USA}

\author{Jason Austermann}
\affiliation{NIST Quantum Sensors Group, 325 Broadway Ave, Boulder, CO 80305, USA}

\author{Andrew O. Bazarko}
\affiliation{Department of Physics, Princeton University, Princeton, NJ 08540, USA}

\author{James A. Beall}
\affiliation{NIST Quantum Sensors Group, 325 Broadway Ave, Boulder, CO 80305, USA}

\author{Sanah Bhimani}
\affiliation{Department of Physics, Yale University, New Haven, CT 06520, USA}

\author{J. Richard Bond}
\affiliation{CITA, University of Toronto, Toronto ON M5S 3H8 Canada}

\author{Grace E. Chesmore}
\affiliation{Department of Physics, University of Chicago, Chicago, IL, 60637, USA}

\author{Steve K. Choi}
\affiliation{Department of Physics, Cornell University, Ithaca, NY 14853, USA}
\affiliation{Department of Astronomy, Cornell University, Ithaca, NY 14853, USA}

\author{Jake Connors}
\affiliation{NIST Quantum Sensors Group, 325 Broadway Ave, Boulder, CO 80305, USA}

\author{Nicholas F. Cothard}
\affiliation{Department of Applied and Engineering Physics, Cornell University, Ithaca, NY 14853, USA}

\author{Mark Devlin}
\affiliation{Department of Physics and Astronomy, University of Pennsylvania, 209 South 33rd Street, Philadelphia, PA 19104, USA}

\author{Simon Dicker}
\affiliation{Department of Physics and Astronomy, University of Pennsylvania, 209 South 33rd Street, Philadelphia, PA 19104, USA}

\author{Bradley Dober}
\affiliation{Department of Physics, University of Colorado Boulder, Department of Physics 390 Boulder, CO 80309, USA}

\author{Cody J. Duell}
\affiliation{Department of Physics, Cornell University, Ithaca, NY 14853, USA}

\author{Shannon M. Duff}
\affiliation{NIST Quantum Sensors Group, 325 Broadway Ave, Boulder, CO 80305, USA}

\author{Rolando Dünner}
\affiliation{Instituto de Astrof\'isica and Centro de Astro-Ingenier\'ia, Facultad de F\'isica, Pontificia Universidad Cat\'olica de Chile, Av. Vicu\~na Mackenna 4860, 7820436, Macul, Santiago, Chile}

\author{Giulio Fabbian}
\affiliation{School of Physics and Astronomy, Cardiff University, The Parade, Cardiff, CF24 3AA, UK}

\author{Nicholas Galitzki}
\affiliation{Department of Physics, University of California San Diego, La Jolla, CA 92093, USA}

\author{Patricio A. Gallardo}
\affiliation{Department of Physics, Cornell University, Ithaca, NY 14853, USA}

\author{Joseph E. Golec}
\affiliation{Department of Physics, University of Chicago, Chicago, IL, 60637, USA}

\author{Saianeesh K. Haridas}
\affiliation{Department of Physics and Astronomy, University of Pennsylvania, 209 South 33rd Street, Philadelphia, PA 19104, USA}

\author{Kathleen Harrington}
\affiliation{Department of Astronomy and Astrophysics, University of Chicago, Chicago, IL, 60637, USA}

\author{Erin Healy}
\affiliation{Department of Physics, Princeton University, Princeton, NJ 08540, USA}

\author{Shuay-Pwu Patty Ho}
\affiliation{Department of Physics, Stanford University, Stanford, CA 94305, USA}

\author{Zachary B. Huber}
\affiliation{Department of Physics, Cornell University, Ithaca, NY 14853, USA}

\author{Johannes Hubmayr}
\affiliation{NIST Quantum Sensors Group, 325 Broadway Ave, Boulder, CO 80305, USA}

\author{Jeffrey Iuliano}
\affiliation{Department of Physics and Astronomy, University of Pennsylvania, 209 South 33rd Street, Philadelphia, PA 19104, USA}

\author{Bradley R. Johnson}
\affiliation{Department of Astronomy, University of Virginia, Charlottesville, VA 22904, USA}

\author{Brian Keating}
\affiliation{Department of Physics, University of California San Diego, La Jolla, CA 92093, USA}

\author{Kenji Kiuchi}
\affiliation{Department of Physics, The University of Tokyo, 7-3-1 Hongo, Bunkyo, Tokyo 113-0033, Japan}

\author{Brian J. Koopman}
\affiliation{Department of Physics, Yale University, New Haven, CT 06520, USA}

\author{Jack Lashner}
\affiliation{Department of Physics and Astronomy, University of Southern California, Los Angeles CA, 90089, USA}

\author{Adrian T. Lee}
\affiliation{Physics Department, University of California, Berkeley, CA 94720, USA}

\author{Yaqiong Li}
\affiliation{Department of Physics, Cornell University, Ithaca, NY 14853, USA}
\affiliation{Kavli Institute at Cornell for Nanoscale Science, Cornell University, Ithaca, NY 14853, USA}

\author{Michele Limon}
\affiliation{Department of Physics and Astronomy, University of Pennsylvania, 209 South 33rd Street, Philadelphia, PA 19104, USA}

\author{Michael Link}
\affiliation{NIST Quantum Sensors Group, 325 Broadway Ave, Boulder, CO 80305, USA}

\author{Tammy J Lucas}
\affiliation{NIST Quantum Sensors Group, 325 Broadway Ave, Boulder, CO 80305, USA}

\author{Heather McCarrick}
\affiliation{Department of Physics, Princeton University, Princeton, NJ 08540, USA}

\author{Jenna Moore}
\affiliation{School of Earth and Space Exploration, Arizona State University, Tempe, AZ 85287, USA}

\author{Federico Nati}
\affiliation{Department of Physics, University of Milano-Bicocca, Milano (MI), Italy}

\author{Laura B. Newburgh}
\affiliation{Department of Physics, Yale University, New Haven, CT 06511, USA}

\author{Michael D. Niemack}
\affiliation{Department of Physics, Cornell University, Ithaca, NY 14853, USA}
\affiliation{Department of Astronomy, Cornell University, Ithaca, NY 14853, USA}
\affiliation{Kavli Institute at Cornell for Nanoscale Science, Cornell University, Ithaca, NY 14853, USA}

\author{Elena Pierpaoli}
\affiliation{Department of Physics and Astronomy, University of Southern California, Los Angeles, CA 90089, USA}

\author{Michael J. Randall}
\affiliation{Department of Physics, University of California San Diego, La Jolla, CA 92093, USA}

\author{Karen Perez Sarmiento}
\affiliation{Department of Physics and Astronomy, University of Pennsylvania, 209 South 33rd Street, Philadelphia, PA 19104, USA}

\author{Lauren J. Saunders}
\affiliation{Department of Physics, Yale University, New Haven, CT 06520, USA}

\author{Joseph Seibert}
\affiliation{Department of Physics, University of California San Diego, La Jolla, CA 92093, USA}

\author{Carlos Sierra}
\affiliation{Department of Physics, University of Chicago, Chicago, IL, 60637, USA}

\author{Rita Sonka}
\affiliation{Department of Physics, Princeton University, Princeton, NJ 08540, USA}

\author{Jacob Spisak}
\affiliation{Department of Physics, University of California San Diego, La Jolla, CA 92093, USA}

\author{Shreya Sutariya}
\affiliation{Department of Physics, University of Chicago, Chicago, IL, 60637, USA}

\author{Osamu Tajima}
\affiliation{Department of Physics, Kyoto University, Kitashirakawa-Oiwakecho, Sakyo-ku, Kyoto 606-8502, Japan}

\author{Grant P. Teply}
\affiliation{Department of Physics, University of California San Diego, La Jolla, CA 92093, USA}

\author{Robert J. Thornton}
\affiliation{Department of Physics, West Chester University of Pennsylvania, West Chester, PA 19383, USA}
\affiliation{Department of Physics and Astronomy, University of Pennsylvania, 209 South 33rd Street, Philadelphia, PA 19104, USA}

\author{Tran Tsan}
\affiliation{Department of Physics, University of California San Diego, La Jolla, CA 92093, USA}

\author{Carole Tucker}
\affiliation{School of Physics and Astronomy, Cardiff University, The Parade, Cardiff, CF24 3AA, UK}

\author{Joel Ullom}
\affiliation{NIST Quantum Sensors Group, 325 Broadway Ave, Boulder, CO 80305, USA}

\author{Eve M. Vavagiakis}
\affiliation{Department of Physics, Cornell University, Ithaca, NY 14853, USA}

\author{Michael R. Vissers}
\affiliation{NIST Quantum Sensors Group, 325 Broadway Ave, Boulder, CO 80305, USA}

\author{Samantha Walker}
\affiliation{Department of Astrophysical and Planetary Sciences, University of Colorado Boulder, Boulder, CO 80309, USA}
\affiliation{NIST Quantum Sensors Group, 325 Broadway Ave, Boulder, CO 80305, USA}

\author{Benjamin Westbrook}
\affiliation{Physics Department, University of California, Berkeley, CA 94720, USA}

\author{Edward J. Wollack}
\affiliation{NASA Goddard Space Flight Center, Greenbelt, MD 20771, USA}

\author{Mario Zannoni}
\affiliation{Department of Physics, University of Milano-Bicocca, Milano (MI), Italy}

\vspace{6mm}




\begin{abstract}

The Simons Observatory (SO) Large Aperture Telescope Receiver (LATR) will be coupled to the Large Aperture Telescope located at an elevation of 5,200\,m on Cerro Toco in Chile.  The resulting instrument will produce arcminute-resolution millimeter-wave maps of half the sky with unprecedented precision.  The LATR is the largest cryogenic millimeter-wave camera built to date with a diameter of \dia{} and a length of \len{}. It cools 1200\,kg of material to 4\,K and 200\,kg to 100\,mk, the operating temperature of the bolometric detectors with bands centered around 27, 39, 93, 145, 225, and 280\,GHz.  Ultimately, the LATR will accommodate 13 40\,cm diameter optics tubes, each with three detector wafers and a total of 62,000 detectors. The LATR design must simultaneously maintain the optical alignment of the system, control stray light, provide cryogenic isolation, limit thermal gradients, and minimize the time to cool the system from room temperature to 100\,mK.  The interplay between these competing factors poses unique challenges.  We discuss the trade studies involved with the design, the final optimization, the construction, and ultimate performance of the system.  

\end{abstract}



\section{Introduction} \label{sec:intro}

Observations of the cosmic microwave background (CMB) are a crucial tools in developing our understanding of the physics of the early universe and testing the standard model of cosmology, $\Lambda$CDM. While satellite missions such as the Cosmic Background Explorer (COBE)~\citep{Smoot1999}, the Wilkinson Microwave Anisotropy Probe (WMAP)~\citep{benn13,hins13}, and the Planck Collaboration~\citep{Planck2020} have produced full-sky microwave maps, ground-based experiments have extended the satellite measurements towards smaller angular scales and lower noise levels. High resolution (${\sim}\,1'$) experiments, such as the Atacama Cosmology Telescope~\citep{Thornton2016} (ACT) and the South Pole Telescope~\citep{Carlstrom_2011, sayre2020} (SPT) have made measurements of both the primordial  power spectrum~\citep{Wu2019} as well as secondary anisotropies such as the thermal and kinematic Sunyaev-Zel'dovich (SZ) effects ~\citep{Hilton2018, Sunyaev1970, Sunyaev1972} and gravitational lensing effects~\citep{Story2015}. Low  resolution experiments (${\sim}\,0.5^\circ$), such as the BICEP and Keck Arrays~\citep{BICEP2014, keck:2018}, the SPIDER~\citep{gual18}, the Atacama B-mode Survey (ABS)~\citep{kusaka/etal:2018}, the POLARBEAR~\citep{POLARBEAR2014, polarbear2020}, and the Cosmology Large Angular Scale Surveyor (CLASS)~\citep{xu/etal:2020c, harr16} aim to improve B-mode polarization measurements at larger angular scales ($\ell\lesssim200$). 

The Simons Observatory~(SO)~\citep{sofc19, gali18} is a next generation CMB experiment consisting of three 0.42\,m Small Aperture Telescopes (SATs)~\citep{ali/etal:2020} and one 6\,m Large Aperture Telescope (LAT)~\citep{pars18} in the Atacama Desert of Chile. The LAT utilizes a crossed-Dragone optical design with a 6\,m primary mirror~\citep{gallardo/etal:2018, gudmundsson/etal:2020}.  The Large Aperture Telescope Receiver (LATR)~\citep{zhu18, orlo18, xu/etal:2020b} can contain up to 13 optics tubes (OTs) and will cover six frequency bands centered around 27, 39, 93, 145, 225, and 280\,GHz with polarized dichroic pixels. The initial survey will use seven optics tubes in the nominal configuration with the remaining space allowing for future upgrades. Each OT contains ${\sim}\,$5000 (depending on the frequency band) polarization sensitive transition edge sensor (TES) detectors~\citep{irwin/hilton:2005} operating below 100\,mK. The LATR is \len{} in length, \dia{} in diameter, and 11\,$\textrm{m}^3$ in volume. It is installed on a bore-sight rotation mount in the LAT receiver cabin where it uses a 6.7$^{\circ}$ field of view out of the $7.8^\circ$ available.

\begin{deluxetable*}{c c c c c c c}

\tablehead{\colhead{Frequency}  & \colhead{FWHM} & \colhead{Baseline} & \colhead{Goal} & \colhead{Frequency} & \colhead{Detector}& \colhead{Optics}\\[-2mm]
\colhead{(GHz)} & \colhead{(arcmin)} & \colhead{($\mu$K-arcmin)} & \colhead{($\mu$K-arcmin)} & \colhead{Bands} & \colhead{Number}& \colhead{Tubes}}
\tablecaption{The SO LAT Survey Specifications\label{tab:LATSensitivity}}

\startdata
    25 & 8.4*& 71 & 52 &\multirow{2}{*}{LF} & 222 & \multirow{2}{*}{1}\\
    39 & 5.4*& 36  & 27 & & 222 &\\
    \hline
    93 & 2.0 & 8.0 &5.8 & \multirow{2}{*}{MF}& 10,320 &\multirow{2}{*}{4} \\
    145 & 1.2 & 10 & 6.3 & & 10,320 & \\
    \hline
    225 & 0.9 & 22 & 15 & \multirow{2}{*}{UHF}&  5,160 &\multirow{2}{*}{2} \\
    280 & 0.8 & 54 & 37 & & 5,160 &\\
    \hline
\enddata
\tablecomments{The SO LAT projected survey sensitivity with $f_{sky} = 0.4$~\citep{sofc19}. This represents the nominal configuration for the first deployment, with seven of the 13 OTs installed (see Section~\ref{subsec:ot_design} for OT design). The quoted detector numbers assume three detector wafers per OT. 
With the use of dichroic pixels, the six targeted bands are split into three categories: low-frequency (LF), medium-frequency (MF), and ultra-high-frequency (UHF).
}
\vspace{-5mm}
\end{deluxetable*}

Over a five year observing period, we plan to use the LAT to survey ${\sim}$\,40\% of the sky with arcminute resolution and a map noise level of ${\sim}$\,6\,$\mu$K-arcmin in the 93 and 145\,GHz bands (see Table~\ref{tab:LATSensitivity}). Overlapping with existing and upcoming optical surveys, the SO will be able to measure cross correlations of the SO reconstructed lensing potential with the Vera C. Rubin Observatory~\citep{LSST__2019} identified galaxies, and provide tighter constraints on the neutrino mass by combining SO data with Dark Energy Spectroscopic Instrument (DESI)~\citep{DESI2016} baryon acoustic oscillation (BAO) information. The LAT will be able to make precise measurements of the small-scale temperature and polarization power spectra, as well as the CMB lensing spectrum~\citep{sofc19}. Additionally, the LAT will detect tens of thousands of clusters using the thermal SZ effect, as well as numerous extra-galactic sources and transient microwave objects~\citep{Naess2020}. There are even prospects for the LAT to detect Oort clouds around nearby stars \cite{Baxter2019, orlo2019} or additional planets in our Solar System~\citep{Baxter2018}. Finally, observations made with the LAT can be used to delens large scale B-mode signals measured with the SATs. A complete discussion of the LAT science can be found in the SO forecast paper~\citep{sofc19}. 

In this paper, we start with a design overview for the LATR in Section~\ref{sec:motivation}. Then, the detailed design of the LATR is discussed in Section~\ref{sec:latr_design}. The LATR testing and validation results are presented in Section~\ref{sec:latr_validation}. Finally, we summarize the lessons for future development in Section~\ref{sec:future_development} and present conclusions in Section~\ref{sec:conclusion}.

\begin{figure*}
  \centerline{
    \includegraphics[width=0.8\linewidth]
{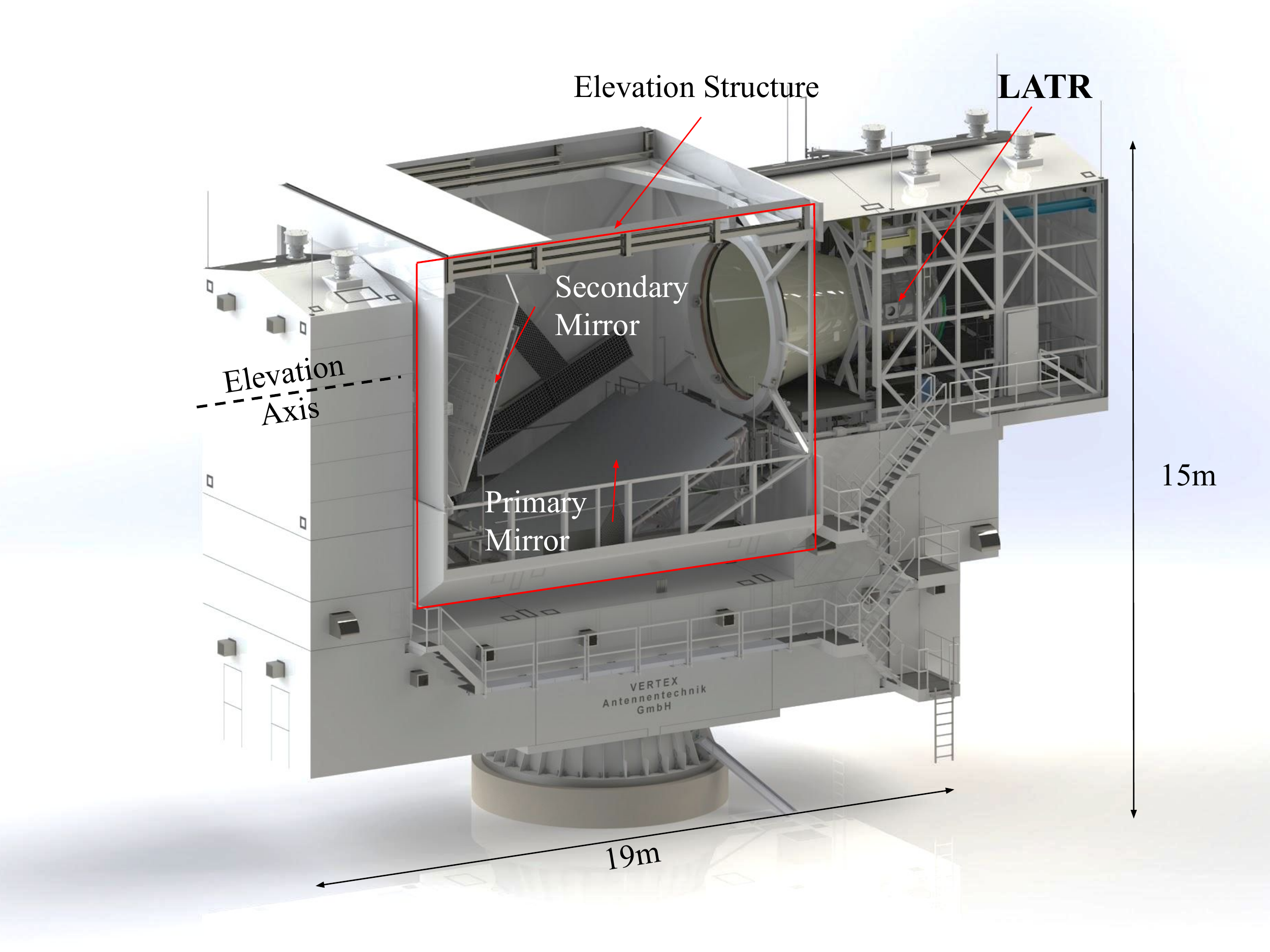}
}
  \caption{The LATR in the LAT. This shows a cutaway rendering of the LAT with the LATR installed in the receiver cabin. The elevation structure is labeled with the 6\,m primary and secondary mirrors inside. As the telescope changes observation elevation by rotating the elevation structure, the LATR co-rotates -- maintaining a consistent orientation between the LATR and the primary and secondary mirrors.\label{fig:LAT_Xsection}}
\end{figure*}

\section{Design Overview} \label{sec:motivation}

\begin{figure}
    \centering
    \includegraphics[width=\linewidth]{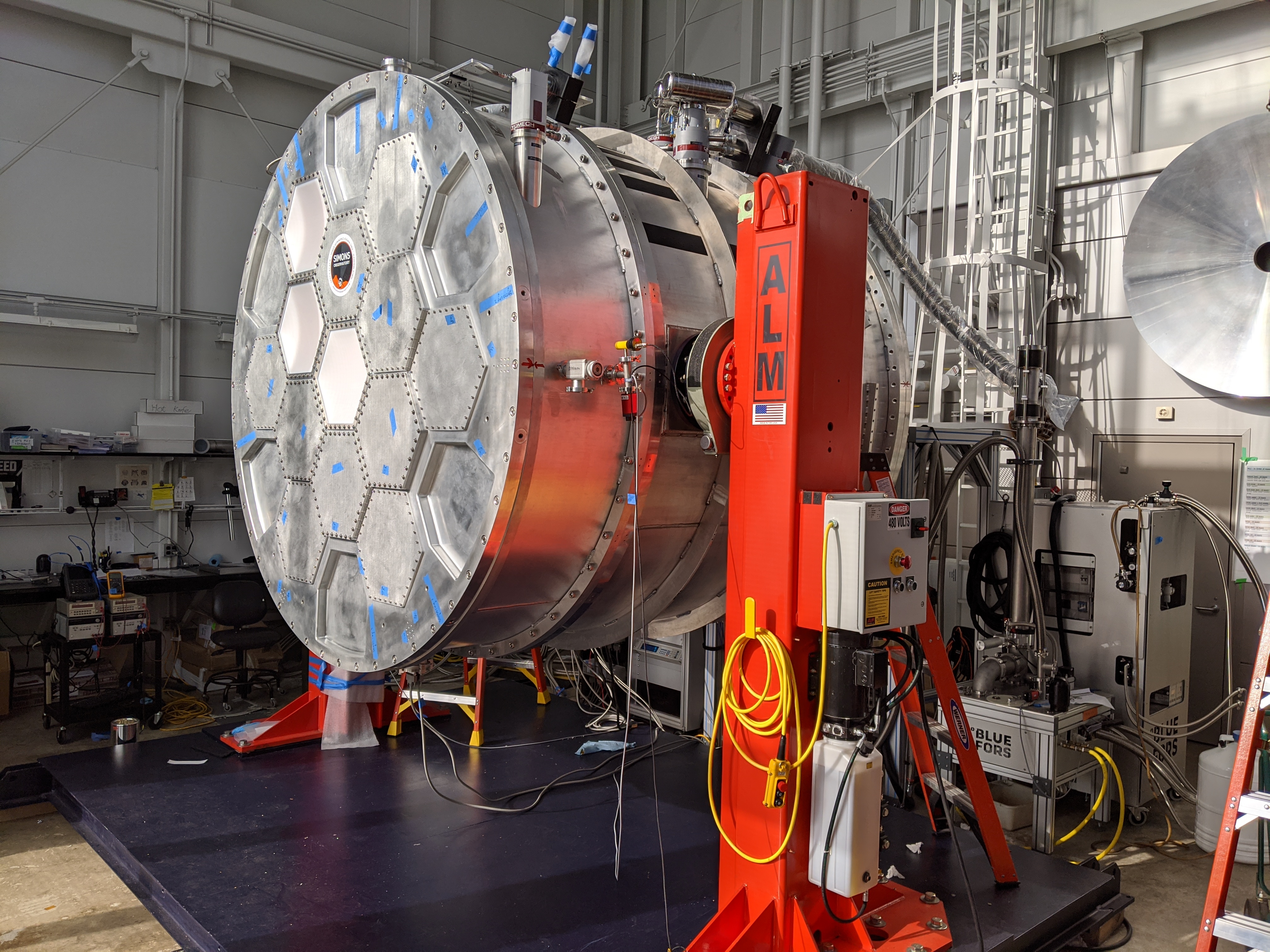}
    \caption{The LATR in the highbay at the University of Pennsylvania. The photo shows the LATR during testing. The \dia{} diameter front plate with 13 hexagonal openings for optics tubes can be seen in the picture. The LATR is held by two arms (red) on a platform (dark blue). The arms are also capable of rotating the LATR with an installed motor.}
    \label{fig:latr_photo}
\end{figure}

The LAT utilizes a cross-Dragone design with a pair of $6$\,m aperture mirrors. The 3-mm-wavelength diffraction-limited focal plane is 1.9\,m in diameter with a $7.8^{\circ}$ field of view~\citep{pars18,Niemack_2016}. While the LATR could have been designed to fill the entire LAT focal plane, we choose to limit the used part of the focal plane to 1.7\,m diameter and a 6.7$^\circ$ field of view out of the $7.8^\circ$ available. The LATR is shown installed in the LAT in Figure~\ref{fig:LAT_Xsection}. A photo of the LATR being tested at the University of Pennsylvania is shown in Figure~\ref{fig:latr_photo}.

Inside the cryostat, cryogenic optics within the OTs re-image the telescope focal plane onto the detector arrays. The detectors are packaged into three universal focal plane modules (UFM), which are cooled to below 100\,mK to improve the detector sensitivity. Metal mesh filters and alumina filters modulate the thermal loading. An advantage of the modularized OT design is that each OT can be optimized for a specific frequency range, as shown in Table~\ref{tab:LATSensitivity}. 




A trade-off study was performed to select the number of OTs for the LATR, with the aim of optimizing sensitivity while accounting for mechanical, optical, and cryogenic constraints~\citep{Hill2018}. Smaller OTs allow for smaller optical elements, which are easier to fabricate. However, one then needs more OTs to fill the same focal plane area, which adds mass and reduces the filling factor due to required spacing for the magnetic shielding, mechanical structures, and cryogenic connections. The additional mass also requires a thicker front plate, which in turn increases the OT separation to avoid clipping the beams exiting the LATR. Balancing the required size of the optical elements with the overhead of additional tubes leads us to an optimum of 13 OTs.


At 5 metric tons with a \dia{} diameter profile, the LATR is the largest sub-Kelvin cryostat to be installed on a steerable telescope to date. The development of this instrument paves the way for future experiments that require large-volume ultra-cold cryostats such as CMB-S4~\citep{s4tb17, s4sb16}.


\section{LATR Design} \label{sec:latr_design}

The design of the LATR vacuum shell is driven by the requirements of being able to withstand atmospheric pressure, the mechanical interfaces associated with mounting it in the LAT, and the desire to minimize the mass that the co-rotating LAT interface would need to support. Based on finite-element analysis (FEA) results, we have adopted a two-piece cylindrical shell design (Figure~\ref{fig:latr_external}). Two aluminum ribs are welded to the back section cylinder for additional strength. The front plate converges to a design with 13 hexagonal openings, reinforced structures and light-weighted pockets on a 6-cm thick aluminum plate (Figure~\ref{fig:latr_external}).

The LATR cryogenic stages mass more than one metric ton. Mechanically supporting this mass with the required stability and optical alignment while minimizing the thermal conductance is a significant challenge.
To solve this, we use thin-walled glass epoxy laminate (G-10) to mechanically support the 80\,K, 40\,K, and 4\,K stages. This design draws on the legacies of both AdvACT~\citep{Thornton2016} and SPIDER~\citep{spid15}. AdvACT used a similar cylindrical G-10 design to mechanically support the 40\,K and 4\,K stages. However, for the LATR, the diameter of such a cylinder would be nearly \dia{}.  When cooled from  300\,K to 40\,K, the diameter of such a structure would contract by 2.5\,cm, resulting in an unacceptable radial stress on the G-10, and significantly weakening the structure.  Inspired by SPIDER, we break up the cylinders by using individual G-10 tabs to support each cryogenic stage.
As detailed in Table~\ref{tab:g10_tab}, the 80\,K stage is supported by 12 G-10 tabs from the front of the vacuum shell; the 40\,K stage is supported by 24 G-10 tabs from the middle of the vacuum shell; and the 4\,K (and colder) stage is supported by another 24 G-10 tabs from the 40\,K stage. The 300-40\,K G-10 tabs can be seen in Figure~\ref{fig:latr_cut}.

The detector arrays and cold optical components at $\le$4\,K (including lenses and filters) are packaged in OTs, which are attached directly to the 4\,K cold plate with relatively easy access. 
The detailed design of the OT is discussed in Section~\ref{subsec:ot_design}. 
Mounting all optical components in the optics path within a single OT makes it simpler to precisely control the relative position of the optical components. The modular design also allows one to easily reconfigure the receiver by swapping out any individual optics tube without impacting the other tubes.
The interior of each OT provides mechanical and thermal isolation for 4\,K, 1\,K, and 100\,mK components. The 4\,K OT stage is attached directly to the 4\,K plate and the 1\,K and 100\,mK stages are cooled via 1\,K and 100\,mK thermal Back-up Structures (BUSs). The 1\,K and 100\,mK thermal BUSs are oxygen-free-high-conductivity (OFHC) copper web structures, which efficiently distribute the cooling power from the dilution refrigerator (DR) to the back of the OTs (Figure~\ref{fig:latr_cut} and Figure~\ref{fig:4K_cavity}).  

The LATR design also incorporates 12 large radial feedthroughs penetrating the 300\,K, 40\,K, and 4\,K stages to accommodate the pulse tubes, the dilution refrigerator, and  cable feedthroughs.  
The radial feedthroughs are designed to remain thermally coupled to the relevant heat shield while avoiding light leaks and allowing for differential motion between the heat shields and 300K mounting plate.
We also paid attention to ensuring the components could be easily installed and removed in keeping with the modular design and future upgrade potential of the 13 optics tube configuration.

\subsection{Mechanical Design} \label{subsec:mechanical_desigh}

\begin{figure}[t]
  \centerline{
    \includegraphics[width=1.1\linewidth]
{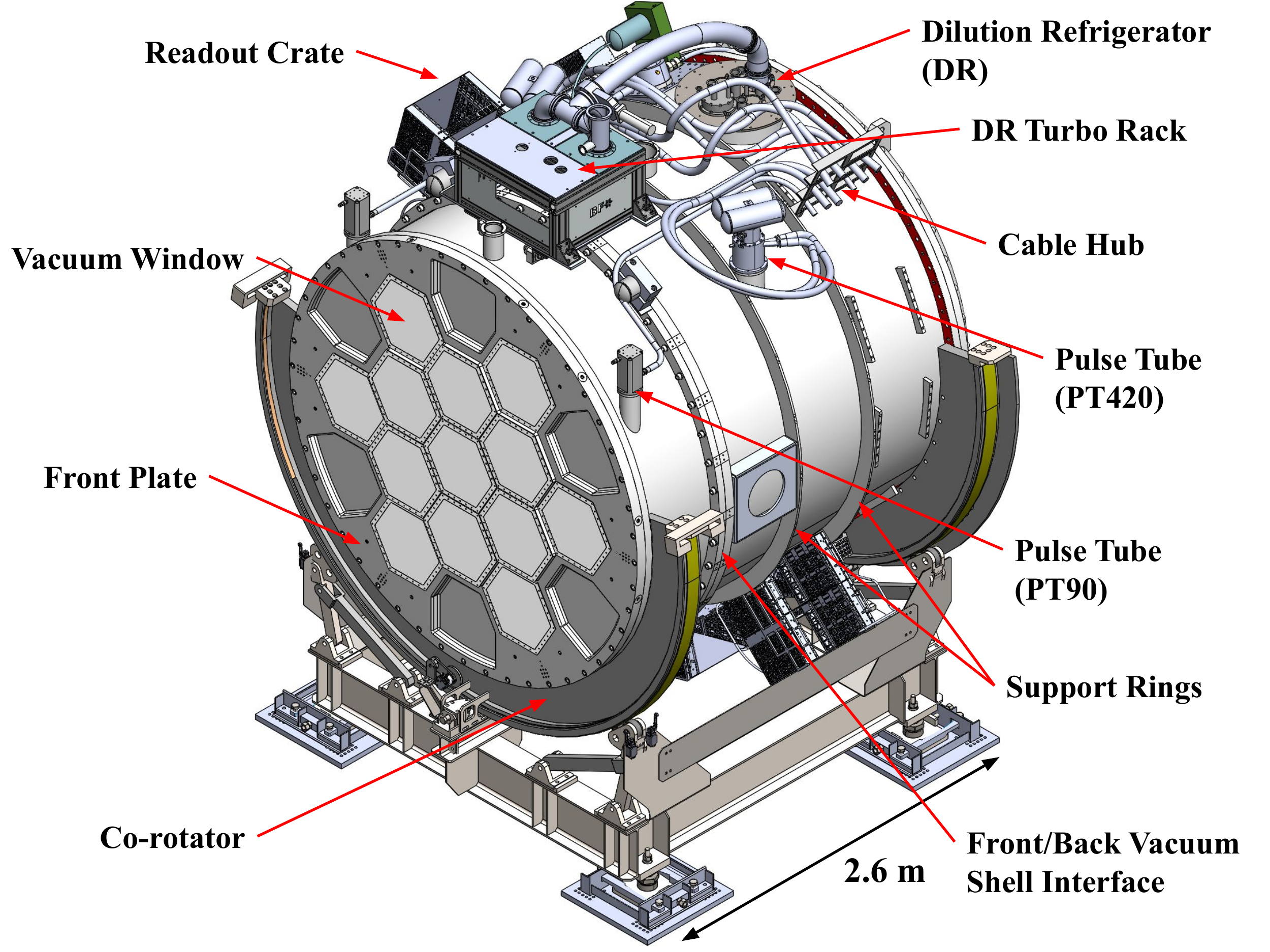}
}
  \caption{LATR external components are shown in this figure. The front plate with 13 hexagonal windows is labeled. The co-rotator at the bottom supports the LATR in the LAT (Figure~\ref{fig:LAT_Xsection}), and allows the LATR to maintain its orientation with respect to the LAT mirrors when the telescope changes elevation. Two of the pulse tube coolers on one side are labeled, including a one-stage 80\,K cooler (PT90) and a two-stage 40\,K/4\,K cooler (PT420). The DR at the back of the LATR is labeled. Also labeled is one of the readout crates around the LATR, which house a number of critical electronics for detector data readout. All hoses and cables travel to the central hub (metal frame on the top right) on the cryostat before going to the cable wrap (Figure~\ref{fig:LAT_Xsection}).
  \label{fig:latr_external}}
\end{figure}

\begin{figure*}
    \centering
    \includegraphics[width=\linewidth]{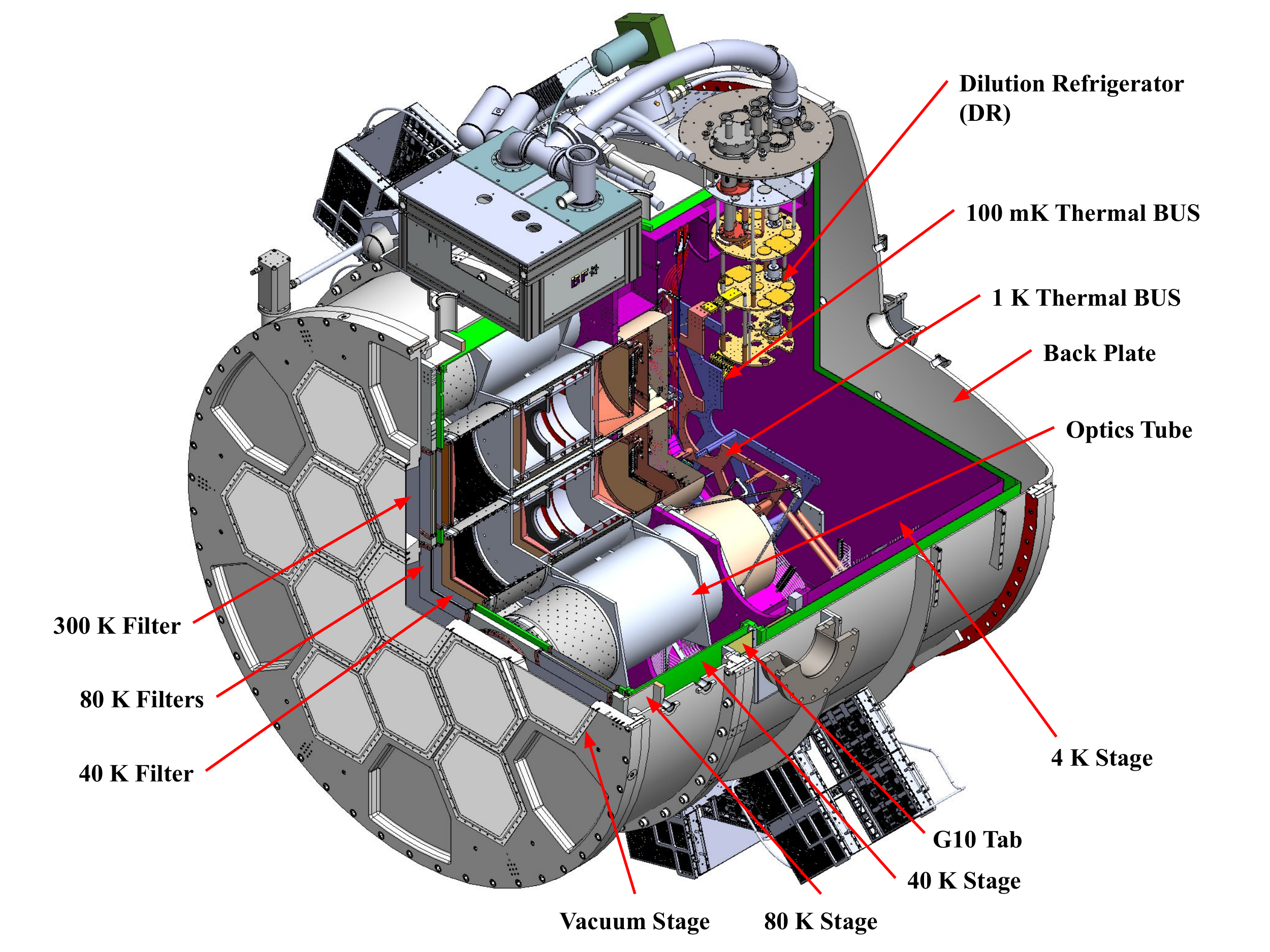}
    \caption{The LATR cut-out to display the internal structures. From the front to the back, the LATR consists of the 300\,K vacuum stage, 80\,K stage (grey), 40\,K stage (green), and 4\,K stage (purple). Section~\ref{sec:latr_design} describes the design of each stage in detail. One of the 300-40\,K G-10 tabs is visible and labeled. Inside the 4\,K cavity, the 1\,K and 100\,mK stages distribute the cooling power from the dilution refrigerator to individual OTs (see Figure~\ref{fig:4K_cavity}). Infrared filters on the cryostat are shown at the 300\,K, 80\,K, and 40\,K stages in front of the OTs. The OTs contain the optical components at $\le$4\,K and the detector arrays. The OT design is discussed in Section~\ref{subsec:ot_design} and its internal structure is displayed in Figure~\ref{fig:OT_cutaway}.}
    \label{fig:latr_cut}
\end{figure*}

The SO LATR mechanical structure is composed of six thermal stages: 300\,K, 80\,K, 40\,K, 4\,K, 1\,K, and 100\,mK, as shown in Figure~\ref{fig:latr_cut}. The vacuum shell, cold plates, and radiation shields were manufactured by Dynavac\footnote{Dynavac, 10 Industrial Park Rd \# 2, Hingham, MA 02043, USA, \url{https://www.dynavac.com/}}.

The 300\,K vacuum shell consists of the front plate, the front shell, the back shell, and the back plate; all constructed of aluminum 6061-T6. The size of the cryostat was chosen to be \dia{} in diameter to accommodate 13 OTs and allow for mechanical supports and thermal isolation/shielding.  

The front plate is a 6\,cm thick flat plate with 13 densely packed hexagonal window cutouts to allow for maximum illumination of the detector arrays.  Optimizing the front plate was one of the most challenging aspects of the LATR design.  Optical and sensitivity requirements warrant maximizing open apertures via closely spaced windows.  However, removing more material drives the plate to be thicker, which eventually leads to a conflict between the diverging optical beam and the window spacing.  Ultimately, the OT spacing was primarily driven by the optimization of the front plate design. The hexagonal windows are tapered to match the diverging beam, leaving as much material as possible to reduce the bending and stress on the front plate.  Alternative designs and materials were considered, including machining the plate with a domed shape, either bowing in or out. These were rejected due to the complexity and expense of machining them. Additionally, doming the front plate would stagger the windows axially; this would require all further optical elements to be axially staggered to maintain consistency of the optical chain, greatly complicating the design of the cold optics.

\begin{figure}
    \centering
    \includegraphics[width=\linewidth]{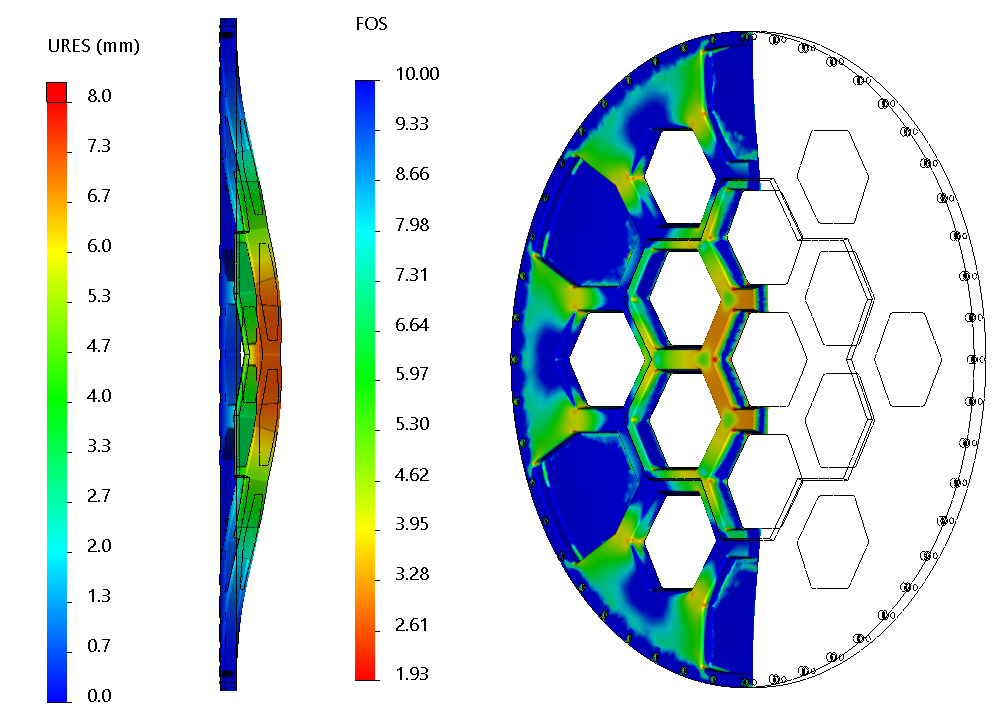}
    \caption{Left: Resultant displacement (URES) plot showing the bowing of the front plate under 1 atmosphere of pressure. The displacement scale is exaggerated 10 times. Right: FoS plot for the vacuum shell under 1 atmospheric pressure at sea-level. The minimum FoS is in the corners of the center window and is due to unphysical stress concentrations from the finite size of the mesh. The actual FoS on the surface of the vessel is $>$\,3.}
    \label{fig:vacuum_fos}
\end{figure}

A key mechanical challenge for the vacuum shell was managing the level of bending of the comparatively narrow struts around the 13 hexagonal optic tube openings in the front plate under atmospheric pressure.
After consultation with an external engineering firm (PVEng\footnote{Pressure Vessel Engineering Limited,
120 Randall Drive, Suite B Waterloo, Ontario, Canada N2V 1C6}), we addressed this concern by increasing the thickness of the front vacuum shell wall and adding stiffening ribs.  Overall the minimum factor of safety (FoS) on the vacuum shell is $>$\,3 at 1 atmosphere of pressure, with the lowest FoS on the front plate. 
Figure~\ref{fig:vacuum_fos} shows the expected deformation of the final front plate design from FEA, magnified by a factor of ten. These results are for sea level atmospheric pressure. While the pressure at the high-elevation site in Chile is only ${\sim}$\,0.5\,bar, the integration testing of the LATR is being done at lower elevations.

Sensitivity requirements drive the windows to be as thin as possible, which has to be balanced against the need to withstand atmospheric pressure. The LATR utilizes 1/8$''$ thick hexagonal windows made of anti-reflection (AR) coated ultra-high molecular weight polyethylene. Each hexagonal window has its own O-ring. A single window has been tested under atmospheric pressure for more than 36 months with 
no indication of reduced structural performance or leakage. The LATR has been tested with seven 1/8$''$ thick windows. 
However, the majority of the in-lab (sea-level) testing has been performed with 1/4” thick windows to remove the risk of a catastrophic failure event that might damage other parts of the receiver.
One double-sided IR blocking filter fabricated by Cardiff University (\cite{ade2006}; section 7) is mounted on the back of the front plate behind each window to reduce the optical loading entering the cryostat. The cold optics design is discussed in detail in Section~\ref{subsubsec:cold_optics}.

The 80\,K stage consists of a 2.1\,m diameter, 2.54\,cm thick circular plate made of aluminum 1100-H14  and a short 80\,K shield made of aluminum 6061-T6. It is designed to intercept most of the optical loading entering the cryostat, reducing the load on the 40\,K stage.  The 80\,K plate contains one double-sided IR blocking filter and one AR coated alumina filter for each OT~\citep{Golec2020}.  The alumina filters act as an IR absorber as well as a prism to bend the off-axis beams back parallel to the long axis of the cryostat~\citep{dick18}. The outside of the 80\,K stage is covered with 30 layers of multi-layer insulation (MLI)\footnote{RUAG Holding AG, Stauffacherstrasse 65, 3000 Bern 22 Switzerland} to reduce the blackbody radiative load coming from the 300\,K shell. The structural support of the 80\,K stage is located at the front of the vacuum shell. Thus, the entire 80\,K stage can be installed or disassembled independently from the rest of the cryostat.

The 40\,K stage consists of two cylinders made of 1100 aluminum and two plates and a ring made of 6061-T6 aluminum. It is suspended directly from the 300\,K vacuum shell, located on the vacuum shell back section (Figure~\ref{fig:latr_cut}). This provides a natural break in the vacuum shell for ease of assembly. There is a thick structural flange on either side of this break that minimizes stress on the vacuum shell itself (Figure~\ref{fig:latr_external}). Like the 80\,K stage, the 40\,K stage is supported by G-10 tabs. Again, we did a thorough FEA validation to verify the structural strength of the G-10 tab support. To reduce loading on the 4\,K stage, the 12.7\,mm 40\,K filter plate at the front contains a third IR blocking filter for each OT. We covered the entirety of the 40\,K assembly with 30 layers of MLI and wrapped each of the G-10 tabs in 20 layers to reduce 300\,K radiation loading. To vent the inside of the 40\,K cavity, evacuation blocks mounted on the back lid of the 40\,K shell were designed to allow air to escape without adding an easy path for light leaks.

The 4\,K stage consists of a 2.08\,m diameter, 2.54\,cm thick circular plate, a 4\,K radiation shield and a thin 4\,K back plate all fabricated from 6061 aluminum. The main 4\,K plate is significantly thicker than the plates for the 40\,K stage since this structure supports the OTs (Section~\ref{subsec:ot_design}). The 4\,K radiation shield has recesses to allow the pulse tube thermal connections to be made without penetrating the 4\,K cavity (Figure~\ref{fig:4K_cavity}). These cutouts allow us space to make thermal contact over a significant area while minimizing the risk of light leaks into the 4\,K cavity, where the 1\,K and the 100\,mK components reside.

\begin{figure}
    \centering
    \includegraphics[width=0.4\textwidth]{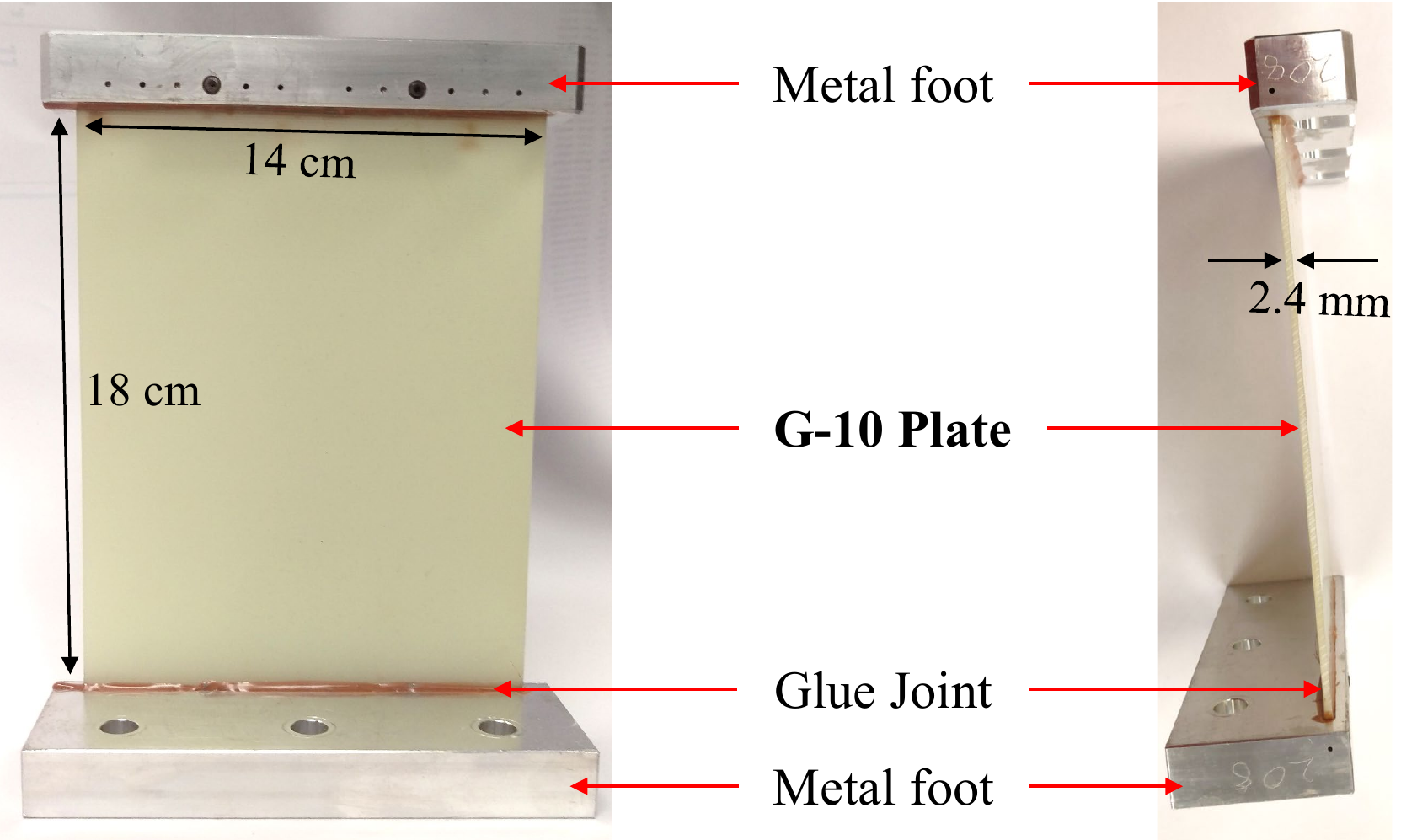}
    \caption{An example of 300-80\,K G-10 tab is shown from two perspectives. Two aluminum 'feet' are glued to G10 plate as mechanical interfaces. Armstrong A-12 was chosen as the glue considering its mechanical and cryogenic performance. Each G-10 tab is serialized as shown in the side-view photo on the right.}
    \label{fig:g10_tab}
\end{figure}

Due to their importance in reducing parasitic thermal load, supporting the mass of the cold stages of the cryostat, and setting the reference location for the 40\,K and 4\,K stages, the G-10 tabs were the subject of significant design focus. The tabs consist of two `feet' connected by a flat sheet of G-10, glued together in a precision jig using Armstrong A-12 epoxy. An example of such tab is shown in Figure~\ref{fig:g10_tab} The tabs can flex radially allowing them to accommodate the high differential thermal contraction (on the order of 2.5\,cm in diameter) between the cold stages and the vacuum shell during cooling.  Extensive FEA was performed to simulate the structural strength of the G-10 tabs~\citep{orlo18}, which determined that the factor of safety for the structural components of the tabs is $>$\,8, excluding the effects of bonding the G-10 tab to the aluminum foot.  A pull test of an assembled tab was performed to determine the strength of the glue bond; it failed via de-adhesion from the aluminum surface at 21\,kN, implying a glue adhesive strength of 590\,MPa. While this force was lower than predicted from the manufacturer listed bond strength, the resulting FoS of 6 still satisifies our requirements.  The geometry and number of tabs are given in Table~\ref{tab:g10_tab}. The G-10 was supplied by Professional Plastics.\footnote{1810 E. Valencia Drive, Fullerton, CA 92831, \url{https://www.professionalplastics.com/}}

\begin{deluxetable}{c c c c c}

\tablehead{\colhead{\multirow{2}{*}{Stage}} & \colhead{Width} & \colhead{Length} & \colhead{Number of} & \colhead{Armstrong A-12 } \\[-3mm]
\colhead{} & \colhead{(mm)} & \colhead{(mm)} & \colhead{Tabs} & \colhead{Glue Area ($\text{cm}^2$)}}
\tablecolumns{5}

\tablecaption{G10 Tab Specifications\label{tab:g10_tab}}

\startdata
	300-80\,K  & 140 & 180 & 12 & 29 \\
	300-40\,K & 160 & 150 & 24 & 35.6 \\
	40-4\,K & 150 & 194.5 & 24 & 33.3\\
\enddata

\tablecomments{The geometry and number of tabs. All tabs are 2.4\,mm thick with different width and length. The G-10 meets NIST G-10 CR process specification and conforms to MIL-I-247682 Type GEE/CR.}
\vspace{-5mm}
\end{deluxetable}

To increase the sensitivity of the TES detectors, all detector components and readout chips are cooled to below 100\,mK~\citep{Mather1984} with a DR manufactured by Bluefors.\footnote{Model LD 400, Arinatie 10, 00370 Helsinki, Finland, \url{https://www.bluefors.com}} To reduce thermal gradients on the 1\,K and 100\,mK stages, we implement two thermal BUSs comprised of two wheel-like structures made of OFHC copper (Figure~\ref{fig:latr_cut} and Figure~\ref{fig:4K_cavity}). Thermal FEA shows that the temperature gradient across both thermal BUSs should be ${\sim}$\,5\,mK, which satisfies our gradient requirements. From the thermal BUS cold finger, individual copper straps, manufactured by TAI,\footnote{Technology Applications, Inc. (TAI), website: \url{https://www.techapps.com/}} extend down to attach to the 1\,K and 100\,mK cold fingers inside each OT. The 1\,K thermal BUS is supported from the 4\,K plate with carbon fiber tripods, while the 100\,mK thermal BUS is supported from the 1\,K thermal BUS via carbon fiber trusses.\footnote{The carbon fiber was sourced from Clearwater Composites, 4429 Venture Ave. Duluth, MN 55811} We choose the twill-ply carbon fiber tubing for these legs as we find that the unidirectional-ply carbon fiber has a tendency to splinter.

\subsection{Cryogenic Design\label{subsec:cryo_design}}

In order to reach the desired detector sensitivity, the LATR cryogenic system must maintain all 62,000 detectors at a stable temperature below 100\,mK while minimizing the in-band optical loading onto the detectors.
The system must also be mechanically rigid, with required tolerances on optical alignment of the elements that are fractions of a millimeter over the \dia{}\,$\times$\,\len{} scale of the system.
Given the LATR needs to cool 1200\,kg of material to 4\,K and 200\,kg to below 100\,mK, it is challenging to design the cryogenic system to cool down the internal structures effectively. 
The 80\,K stage uses two PT90 pulse tubes while the 40\,K and 4\,K stages are cooled by two PT420 pulse tubes supplied by Cryomech\footnote{113 Falso Dr, Syracuse, NY 13211, \url{https://www.cryomech.com}}.  The 1\,K and 100\,mK stages are cooled by the DR, which is backed by an additional PT420 pulse tube, supplementing the cooling power of the main cryostat.  As a design contingency, the LATR can accomodate an additional PT90 and PT420 pulse tube cooler for the 80\,K and 40\,K/4\,K stages respectively.   However, these additional coolers are not required based on our cryogenic validation tests.

Managing and minimizing the thermal loads to meet the cryogenic performance specifications drives almost every aspect of the thermal design.
The base temperature of the pulse tube is a function of the total thermal load on the system. Thermal load on each stage includes conductive load from supporting structures, radiative load from the warmer stages and incoming light, conductive load from cables, and cryogenic electrical components.  Estimates of all of these were included in thermal models, summarized in Table~\ref{tab:latr_loading}.

The large \dia{}\,$\times$\,\len{} scale of the receiver poses its own challenges to the cryogenic design. Reaching the target temperatures at the cold head of the pulse tube coolers or dilution refrigerator is a necessary but not sufficient condition. The thermal design must also include sufficient thermal conductivity within each temperature stage to keep the thermal gradient below the required level. 
For example, the farthest point on the 40\,K filter plate is ${\sim}$\,2\,m from the nearest pulse tube with five mechanical joints in the thermal path (Figure~\ref{fig:latr_cut}).  
If the thermal link is weak, a small amount of power can lead to a large temperature gradient along the path, resulting in an elevated optical filter temperature, without substantively affecting the pulse tube base temperature.
In most cases, the thermal path follows the mechanical structure. However, there are exceptions. For example, 6061 aluminum is used for the mechanical structure of the 4\,K cold plate to meet the optical alignment requirements, and supplemented by the softer but more conductive 1100 aluminum to meet the thermal requirements. In another case, flexible thermal strips connect the coolers and cryogenic stages to transfer heat wile allowing for the differential contraction between the two.

Finally, it is important to minimize temperature gradient during cooldown. 
Bringing the system from 300\,K to the target temperature requires removing $350$ million Joules of energy from the system.  Gradients are important because the pulse tube cooling efficiency is a very steep function of the cold head temperature. 
The PT420 second stage can remove up to 225\,W at 300\,K, but only 2\,W at 4\,K.  
Therefore, the most efficient use of the pulse tube during cooling is achieved when thermal gradients are minimized across the cryostat, while pulse tube cold head temperature is maximized. 
A large temperature gradient between the cold head and stage means the pulse tube temperature is well below the rest of the stage, and thus the pulse tube can remove less power than it would if the entire system was at the same temperature.
A model of the material properties as a function of temperature is included in the design to ensure sufficient thermal conduction throughout the relevant parts of the LATR during cooldown.  The cooldown predictions are shown in Section~\ref{subsubsec:cooldown_simulations}, and the measured cooldown times are presented in Section~\ref{subsec:cryo_validation}.

\subsubsection{Thermal Modeling}
\label{subsubsec:thermal_modeling}
Thermal modeling included a careful accounting of all sources of loading at each stage. 
To compute the thermal loading from supports, we first compiled a library of materials and their thermal conductivities at cryogenic temperatures. We  obtained measurements from the National Institute for Standards and Technology (NIST) cryogenic material properties database~\citep{nist_database, marq2002} and from Adam Woodcraft's low temperature material database~\citep{Woodcraft2009}. The total conductive load for a support is the integrated thermal conductivity between the temperatures on the high and low ends. The conductivity of the cables were computed in the same way, with the conductivity of the coax cables provided by Coax Co.\footnote{COAX CO., LTD. 2-31 Misuzugaoka, Aoba-ku, Yokohama-shi, Kanagawa 225-0016 Japan, \url{http://www.coax.co.jp/en/}}

The thermal conduction due to radiation on surfaces covered in MLI was computed using the Lockhead equation~\citep{keller1974}:

\begin{equation}  
\label{eqn:lockhead}
\begin{split}
    \text{q}_{\text{tot}} &= \frac{\text{C}_{\text{c}}\text{N}^{2.56}\text{T}_{\text{m}}}{\text{n}}(\text{T}_{\text{h}}-\text{T}_{\text{c}})\\
    &+\frac{\text{C}_{\text{r}}\epsilon_0}{\text{n}}(\text{T}_{\text{h}}^{4.67}-\text{T}_{\text{c}}^{4.67}),
\end{split}
\end{equation}
which accounts for both the radiative load between layers of the MLI blanket (right hand side) and the conductive loading through the layers of the blanket (left hand side). In Equation~\ref{eqn:lockhead}, $\text{q}_{\text{tot}}$ is the total thermal load on the MLI, $\text{C}_{\text{c}} = 8.95\times 10^{-5}$ is a numerical constant defining the MLI conductive heat transfer, N is the MLI layer density in layers per centimeter, $\text{T}_{\text{m}}$ is the mean MLI temperature, taken to be $\text{T}_{\text{m}} =\frac{\text{T}_{\text{h}}-\text{T}_{\text{c}}}{2}$, n is the number of MLI layers, $\text{T}_{\text{h}}$ is the hot-side temperature in Kelvin, $\text{T}_{\text{c}}$ is the cold-side temperature in Kelvin, $\text{C}_{\text{r}} = 5.39\times 10^{-7}$ is a numerical constant defining the MLI radiative heat transfer, and $\epsilon_0=0.031$ is the MLI emissivity~\citep{keller1974}.  


\begin{deluxetable*}{c c c c c c c c c c}

\tablehead{\colhead{Stage} & \colhead{Support} & \colhead{Cabling} & \colhead{Radiative} & \colhead{Optical} & \colhead{LNAs} & \colhead{Attenuation} & \colhead{Arrays} & \colhead{Total} & \colhead{Available Power}\\[-2mm]
\colhead{(K)} & \colhead{(W)} & \colhead{(W)} & \colhead{(W)} & \colhead{(W)} & \colhead{(W)} & \colhead{(W)} & \colhead{(W)} & \colhead{(W)} & \colhead{(W)}}

\tablecaption{LATR Thermal Loading Estimates\label{tab:latr_loading}}

\startdata
    80    & 3.17 & 0.0261    & 6.59 & 50.7 & N/A   & N/A   & N/A   & 60.5 & 180 \\
    40   & 9.51 & 14.7 & 25.5 & 0.0256\ & 5.8\ & 0.00958 & N/A   & 55.5 & 110 \\
    4    & 0.313 & 0.776 & 0.197 & 0.359 & 0.576 & 0.00157 & N/A   & 2.22  & 4.00 \\
    1    & 2.04$\text{x10}^{-3}$ & 0.326$\text{x10}^{-3}$  & 5.47$\text{x10}^{-6}$ & 0.378$\text{x10}^{-3}$ & N/A   & 0.111$\text{x10}^{-3}$ & N/A   & 2.85$\text{x10}^{-3}$ & 25.0$\text{x10}^{-3}$ \\
    0.1  & 43.6$\text{x10}^{-6}$& 8.33$\text{x10}^{-6}$ & 0.405$\text{x10}^{-6}$ & 4.06$\text{x10}^{-6}$ & N/A   & N/A   & 10.1$\text{x10}^{-6}$ & 66.5$\text{x10}^{-6}$ & 500$\text{x10}^{-6}$ \\
\enddata

\tablecomments{Loading estimates for each temperature stage of the LATR split by source. The provided load estimates are for 13 OTs. The cooling power at 80\,K is supplied by two PT90 coolers, the power at 40\,K and 4\,K is supplied by two PT420 coolers, the power at 1\,K by the DR still stage, and the 100\,mK by the DR mixing chamber stage.}
\vspace{-5mm}
\end{deluxetable*}

To calculate the radiative and optical loading throughout the LATR, we created a custom Python package that models the thermal performance of all filter elements in an ideal radiative environment. The code estimates the total power emitted and absorbed at each temperature stage by performing a radiative transfer calculation using numerical ray optics for individual optics tubes. The circularly symmetrized geometry, the spectral properties of the filters, and the emissivity of temperature stage walls are used as inputs to this model. An additional output of this simulation is a set of radial temperature profiles for all filter elements (computed using temperature-dependent thermal conductivities of the filter materials). The resultant thermal loads for each stage are shown in Table~\ref{tab:latr_loading}.

Power dissipation from electrical components such as the detector readout low noise amplifiers (LNAs) and TES biasing was also included in the thermal model. The LNA power dissipation was computed by multiplying the LNA drain current by its drain voltage, using the nominal current and voltage from the specification sheet.

We used the COMSOL\footnote{COMSOL, Inc., 100 District Avenue, Burlington, MA 01803, USA, \url{https://www.comsol.com}} software suite to estimate thermal gradients across our 80\,K, 40\,K, 4\,K, 1\,K, and 100\,mK stages. Our simulations used the computer aided design (CAD) model of the relevant thermal stage, with some simplifications that only minimally impact the accuracy of the simulations -- for example, suppressing screw holes. Using the material library we developed for the thermal model we applied a material to each part, specifying its thermal conductivity as a function of temperature. Thermal loads were distributed throughout the model. To include the effect of the relevant cryocooler, a measured load curve of that cryocooler as a temperature dependant negative heat flux is applied to the cold head of the cryocooler. 
These simulations allowed us to identify which areas were cryogenically critical, leading us to make those areas thicker and more conductive. 
The simulations showed that to reduce the gradients throughout the stages, the radiation shields play a crucial role in conducting the cooling power besides shielding radiation. As a result of this for the 40\,K stage that relies heavily on the shield to cool the filter plate, we used Al1100 whose thermal conductivity is significantly higher that Al6061. A summary of our predicted gradients and the LATR performance during validation and testing can be found in Section~\ref{subsec:cryo_validation}.


		
\begin{figure}[t]
  \centerline{
    \includegraphics[width=\linewidth]{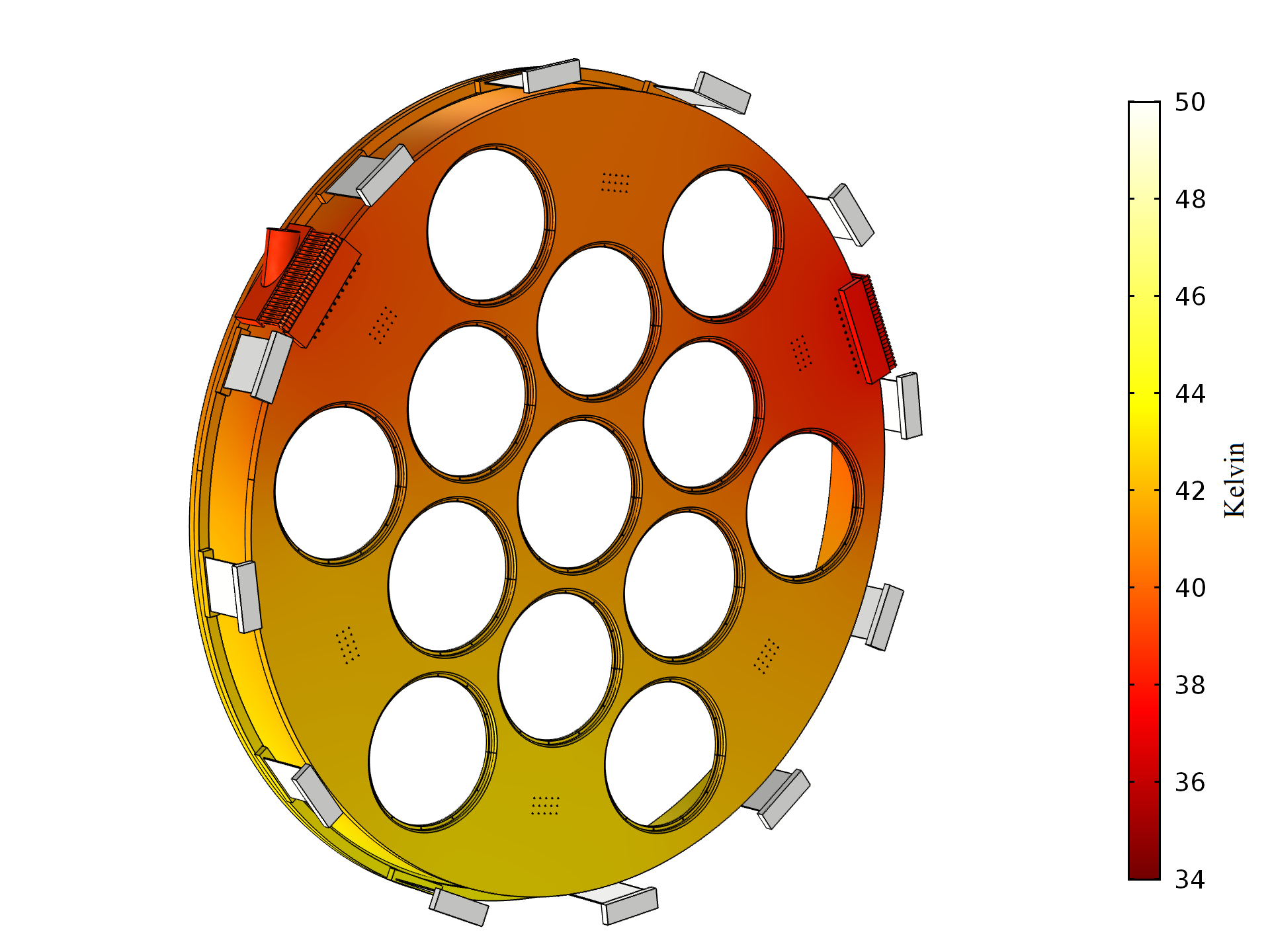}}
  \caption{A thermal simulation of the 80\,K stage. This simulation uses the individually-calibrated PT90 load curves. It was performed to emulate a dark thermal validation run with the windows closed, hence there is no optical load applied. The conductive load through the G10 tabs was included, as is radiative heating from the 300\,K stage. The color map shows the distribution of the thermal gradient with the lowest temperature around the two PT90 thermal straps and the highest temperature at the bottom of the plate. This is one example of the thermal simulations we conducted for all the temperature stages.
    \label{fig:thermal_80K}}
\end{figure}

During the lab testing phase, we performed additional simulations in conjunction with our cryogenic validation. For these simulations, we would mimic a given validation test setup, including whichever components were installed and using the measured cryocooler load-curves. We would then compare the gradients observed in the simulations to those observed in our validation testing. One major difficulty in the thermal validation of our cryostat at all temperature stages was that we were only able to put a limited number of thermometers on each stage, making it difficult for us to identify the source of un-accounted-for loading. This combined with the very long turn around time of our cryostat meant that we risked spending upwards of a month tracking down heat leaks. Given a proposed load, we would add that load in the simulation and compare the resulting temperature gradients to those observed. While the result of the simulation is not precise, missing among other things contact resistance, it is accurate enough to provide a sanity check. An example simulation is shown in Figure~\ref{fig:thermal_80K}. \\

\subsubsection{Cooldown Simulations}
\label{subsubsec:cooldown_simulations}

With a cryostat the size of the LATR, the main challenge for a relatively fast cooldown is distributing the available cooling power and minimizing the gradients during the cooldown. At room temperature, the PT420s each provide up to 450\,W of cooling power for the first stage and 225\,W for the second stage.  If all of the cryogenic material in the LATR were isothermal to the pulse tube heads, the cooldown time would be 5.3 days. However, gradients across the temperature stages reduce the pulse tube heads temperatures, and hence their cooling 
power, so that a realistic simulation is required to estimate the true cooldown time.

\begin{figure}
    \centering
    \includegraphics[width=8.5cm]{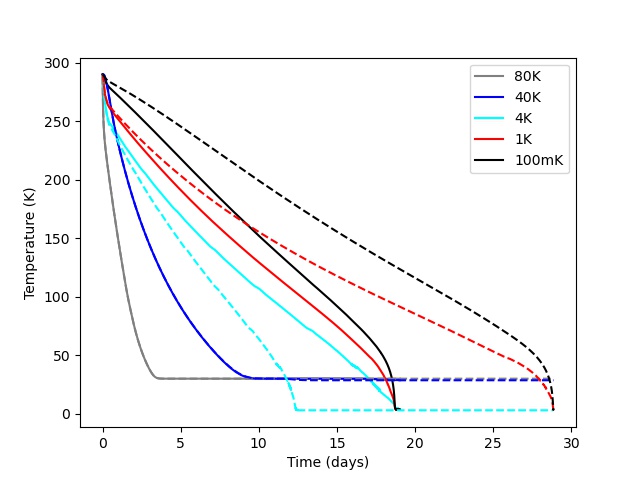} 
    \caption{Simulated cooldown curve for a fully populated LATR with 13 OTs, showing each temperature stage. The solid lines show the result with mechanical heat switches connecting the 4\,K stage with the 1\,K and 100\,mK stages while the dashed lines show the result without heat switches. The heat switches significantly reduce expected cooling time.}
    \label{fig:LATRcooldownsimulation}
\end{figure}

We developed a code based on a finite difference method as described in~\citet{coppi2018}. The results for the LATR with 13 OTs are presented in Figure~\ref{fig:LATRcooldownsimulation}. The warmer stages -- 80\,K, 40\,K and the 4\,K -- cool relatively quickly, in less than 15 days. However, initial simulations showed that it could take up to 28 days for the coldest stages -- 1\,K stage and 100\,mK stage -- to reach 4\,K.\footnote{Mixture condensation, the process in which the DR reaches its base temeprature of 100\,mK proceeds in a few hours once the DR has reached 4\,K.} A detailed  cooldown time measurement with 1 OT and 3 filter sets is presented in Section \ref{subsec:cryo_validation}.

 To reduce cooldown time, we installed a nitrogen heat pipe and two mechanical heat switches. We studied multiple options for accelerating the cooldown process, such as using an external cooler to force cold gas circulation in an internal vacuum chamber~\citep{Alduino2019}, liquid nitrogen flow through a plumbing system~\citep{Lizon2010}, nitrogen heat pipes connecting 4\,K stage to the colder stages~\citep{Duband2018}, and the use of mechanical heat switches from 4\,K stage to the colder stages. The first two options were discarded due to mechanical design complications and the risk of developing large temperature gradients in the cryostat that would induce thermal stresses on critical components such as lenses and filters. Moreover, as stated above, the LATR \emph{has} sufficient cooling power, but requires optimization of its distribution. 
 We chose to apply the remaining two options because they are relatively easy to implement from a mechanical design point of view. Nitrogen heat pipes are intrinsically passive elements with high thermal conductivity between 63\,K and 110\,K and close to zero outside this range. A nitrogen heat pipe was custom installed on the DR by Bluefors. Mechanical heat switches have very high thermal conductivity when closed and zero when open, but require some activation mechanism. Mechanical heat switches manufactured by Entropy\footnote{Entropy GmbH, \url{http://www.entropy-cryogenics.com/}} were chosen for the LATR. These heat switches have a nominal conductance of 0.5\,W/K at 4\,K when closed. Figure~\ref{fig:4K_cavity} shows a photo of the heat switches installed on the 4\,K plate. One switch connects the 4K and 1K stages while the other connects the 4K and 100 mK stages. With these heat switches, a fully-equipped LATR (with 13 OTs) would cool in around 18 days according to our simulations.

\subsection{\label{subsec:ot_design}Optics Tubes}

\begin{figure*}[t]
  \centerline{
    \includegraphics[width=\linewidth]
{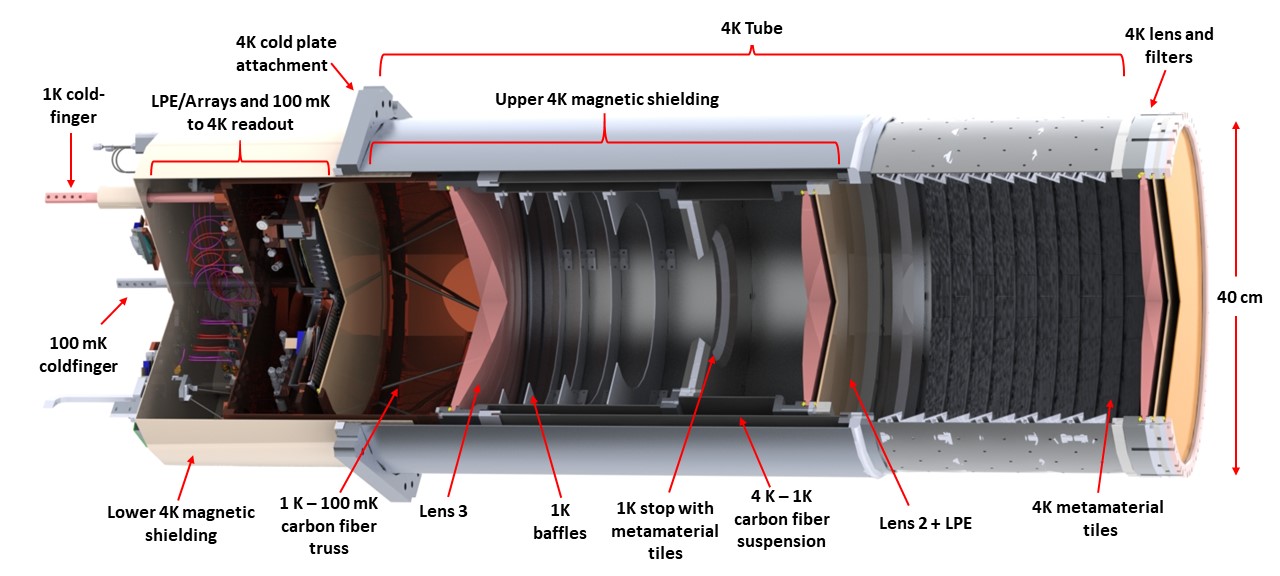}
}
  \caption{A rendering showing the major parts of one OT. Light rays enter from the right through the 40\,cm diameter 4\,K lens and filters. After the metamaterial tile section, light rays are refracted by Lens\,2. Then the incoming light rays are truncated by the metamaterial tile covered Lyot stop before going in to the ring baffle section. Finally, the light rays are focused by Lens\,3 before reaching the detector arrays. The detector arrays and their supporting components are presented in Figure~\ref{fig:OT_Array_fig} with more details.
    \label{fig:OT_cutaway}}
\end{figure*}

The cryostat can accommodate a total of 13 optical chains, each with a dedicated set of detectors (see Table~\ref{tab:LATSensitivity}). Each optical chain consists of elements that are mounted on either the cryostat cold plates (300\,K, 80\,K, and 40\,K) or, for the colder stages, in a large, self-contained OT (Figure~\ref{fig:OT_cutaway}) that has components at 4\,K, 1\,K, and 100\,mK. Each tube is roughly 40\,cm in diameter and 130\,cm long and is designed to be removable as a single unit from the rear of the cryostat. 
As detailed in Table~\ref{tab:LATSensitivity}, the initial seven optics tubes include 1 in the LF bands, 4 in the MF bands, and 2 in the UHF bands.
A cross-section of an MF tube is shown in  Figure~\ref{fig:OT_cutaway}. The field of view for each OT is 1.3$^{\circ}$ in diameter, which is only partially filled by the three detector wafers.

\subsubsection{\label{subsubsec:cold_optics}Cold Optics}
We selected a refractive cold optics design for its compactness. Inside each OT, three silicon lenses re-image the telescope focal plane onto a set of three hexagonal detector arrays \citep{dick18}. Silicon is selected as the lens material due to its low loss and the outstanding performance of developed metamaterial AR coatings~\citep{Golec2020, Coughlin_2018, datta/etal:2013}.  The upper edge of each frequency channel is set by a set of low-pass edge (LPE) filters, which use a capacitive mesh design~\citep{ade2006} to set a range of cut-off frequencies (e.g., 12.5\,cm$^{-1}$ to 6.2\,cm$^{-1}$ for MF), are incorporated to reduce optical loading on different temperature stages. The cut-offs are also chosen to suppress out-of-band leaks from other filters. The LPE filters are complemented by infrared blocking filters at warmer stages~\citep{Tucker2006} to reduce the thermal loads on the colder stages. The designs implemented for the SO incorporate two metal mesh patterns printed onto each face of a polypropylene substrate to form double-sided IR blockers. The capacitive square copper patterns have a 15\,$\mu$m period; the polypropylene substrate is thin (4\,$\mu$m) and absorbs very little of the in-band radiation.  Thus each device acts as a basic low-pass filter, which is optimized for high reflectivity of near-IR radiation with virtually no attenuation of the science bands.  Successive filters are placed between 300\,K and 4\,K to reflect the IR radiation emitted by these stages and to protect the thermal environment at the 1\,K and 0.1\,K stages.  Combined, the optical elements consist of: a 3.1\,mm thick AR coated window and an IR blocking filter at 300\,K;\ an IR blocking filter and alumina absorbing filter at 80\,K; an IR blocking filter at 40\,K; an IR blocking filter, LPE filter, and the first lens at 4\,K; an LPE filter, the second lens, the Lyot stop, and third lens at 1\,K; and the final LPE filter and detector arrays at 100\,mK. See Figure~\ref{fig:OT_cutaway} for the relative locations of optical elements inside the OT.    
 
In this optics tube design, it is worth highlighting part of the 80K design that makes it possible. The alumina filter at 80K is wedge-shaped (except for the central OT) so that it acts as a prism in addition to acting as an IR blocking filter. This prism allows all OTs to be coaxial with the cryostat, which allows
for a much more efficient use of space and greatly simplifies the installation/removal of the OTs. Each alumina filter has a notch machined into it for setting the proper angular orientation.  We are pursuing several parallel AR coating techniques for this element to optimize cost versus performance.  The baseline is to use metamaterial coatings on alumina~\citep{Golec2020}.
 
The close-packed OTs prevent the use of traditional light baffling between 4\,K filter and Lens 2, since there is little space between the walls of the upper (4\,K) OT and the incoming beam. Since stray light is a major concern, significant effort was dedicated toward fabricating novel, low-profile metamaterial microwave absorbers~\citep{xu/etal:2020, gudmundsson/etal:2020}. The goal was a design that had maximum absorptivity and minimum reflection, was capable of cooling down to cryogenic temperatures while maintaining mechanical integrity, and was relatively easy to install. The solution was injection-molded, carbon-loaded plastic tiles that create a gradient index AR coating~\citep{xu/etal:2020}. About 240 holes were laser cut into the upper tube, allowing each tile to be screwed into place. A flat version of the tile was designed to attach to both sides of the 1\,K Lyot stop. The region in between the stop and Lens 3 was baffled with standard ring baffles covered with a mixture of Stycast 2850 FT, coarse carbon powder, and fine carbon powder, because it has more radial clearance and is less critical for stray light. The 1\,K radiation shield surrounding the detector arrays was blackened in the same manner.

\subsubsection{Mechanical Support Structures}
The windows and all filters are mounted using aluminum clamps (except for the 100\,mK LPE filter, which uses a copper clamp).  The LPE filter clamps are axially spring-loaded to minimize the possibility of delamination due to the shear forces involved during large (radial) differential thermal contraction (${\sim}$\,6\,mm) between the aluminum mounts and the polypropylene filters.  The spring is a commercial spiral beryllium copper structure.\footnote{http://www.spira-emi.com/} The lens mounts contain both an axial spring and a radial spring.  The axial spring ensures firm thermal contact between the lens and mount without risking cracking the brittle silicon.  The radial spring ensures the lens is centered at operating temperature by pushing the lens against two opposing hard points.

Each OT consists of a large 4\,K cylindrical structure containing the 1\,K and 100\,mK components.  A 4\,K A4K\footnote{Amuneal 4K material, website: \url{https://www.amuneal.com/magnetic-shielding/magnetic-shielding-materials}} magnetic shield lines the outside wall of the 4\,K tube, extending from the location of the second lens to the rear end of the OT. The 4\,K and 1\,K tubular sections were fabricated from aluminum 1100-H14 sheet to reduce thermal gradients along their lengths; an improvement according to simulations of 2\,K as compared to Al~6061) . The sheets were welded to Al~6061 O-temper flanges. A final machining step is done to attain the assembly specifications. The 1\,K components  are supported from the 4\,K tube by a custom-made carbon fiber tube from Clearwater Composites and Van Dijk.\footnote{Van Dijk Pultrusion Products (DPP BV). Website: \url{https://www.dpp-pultrusion.com/en/the-company/}} A 1\,K radiation shield is installed around the 100\,mK components.  Because the shield serves the dual purpose of supporting and cooling readout components, it was fabricated out of OFHC copper.  A carbon fiber truss attaches to the rear of the 1\,K section and supports all 100\,mK components. Mechanical loading tests verified that the carbon fiber structures were able to support at least five times the expected operating loads.  Simulations show the lowest vibrational frequency is above 50\,Hz.

\begin{figure*}[t]
  \centerline{
    \includegraphics[width=6.5in]
{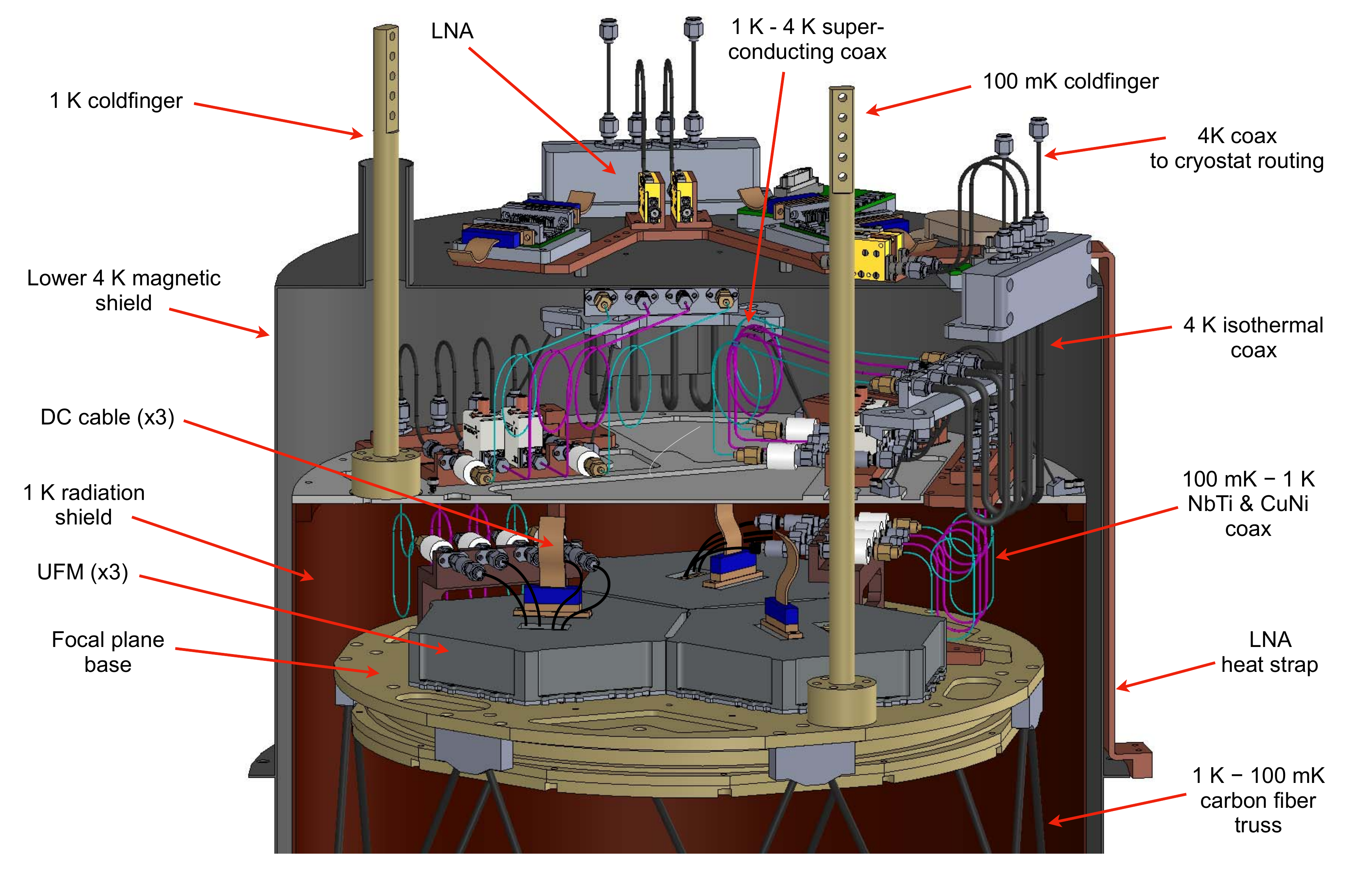}
}
\caption{OT mechanical and readout components on the 4\,K, 1\,K, and 100\,mK stages. This design shows the cable setup for a multiplexing factor of 1,000, where each UFM is read out by two pairs of RF lines. In addition, one DC ribbon cable per UFM provides the detector biases and flux ramps. 
\label{fig:OT_Array_fig}}
\end{figure*}

\subsubsection{\label{subsubsec:ot_det_array_readout}Detector Arrays and Readout}
The SO uses two types of  dual-polarization dual-frequency TES bolometer arrays. For MF and UHF frequencies, the TES bolometers are coupled to the optics with orthomode transducers  feeding monolithic feedhorn arrays, based on their well-tested performance in the ACT experiment~\citep{Choi_2020, Henderson2016, Duff2016}.  For LF frequencies, sinuous antennas with  lenslets couple the bolometers, based on the design successfully used for POLARBEAR~\citep{Suzuki2016} and the South Pole Telescope~\citep{pan/etal:2018}.  For both architectures, each TES array is hexagon-shaped and fabricated from wafers 150\,mm in diameter.  A detector array, including the light-coupling mechanism, TES bolometers, 100\,mK readout  architecture, and associated magnetic shielding are packaged into a single UFM~\citep{McCarrick2021}. There are three UFMs per OT. The UHF and MF OTs contain a total of 1,290 pixels that couple to 5,160 optically active TESes. An LF tube has 111 pixels that couple to 444 optically active TESs. Details of the UFM design are described in~\citet{Li2020}. 

Due to space limitations, one of the most significant challenges of the OT design was the routing of the readout cabling from the three UFMs at 100\,mK to the 4\,K components on the back of the OT magnetic shielding (Figure~\ref{fig:OT_Array_fig}). The SO uses microwave multiplexing technology to read out the detectors~\citep{Dober2021}. For the target multiplexing factor of 1,000, each OT needs 12 radio frequency (RF) coaxial cables to read out ${\sim}\,5000$ detectors and 3 direct current (DC) ribbon cables for detector/amplifier biases and flux ramps~\citep{Mates2012}. 

To simplify the routing between isothermal components, hand-formable 3.58\,mm copper coaxial cables were used.\footnote{Mini-Circuits https://www.minicircuits.com/ }
For connecting OTs readout components at different temperatures (4\,K\,--\,1\,K and 1\,K\,--\, 100\,mK), low thermal conductivity semi-rigid cables were employed.  
CuproNickel (CuNi) cables with 0.86\,mm diameter are used for RF$_{\rm in}$ to control the attenuation at each step to a colder temperature stage, as well as limit thermal loading. For the RF$_{\rm out}$ lines, superconducting 1.19\,mm Niobium Titanium (NbTi) cables maximize signal-to-noise, yet still have good thermal isolation properties.  Both types of semi-rigid cables are bent into loops for strain relief.  DC blocks are implemented for additional electrical/thermal isolation.  Attenuators on the input lines guarantee that the correct power level is delivered to the resonators, the multiplexing components in the array. 
Each 4\,K RF output line has a LNA mounted on the back of the magnetic shield.  To reduce temperature rise due to the power generated by the LNAs (${\sim}$\,5\,mW each), the LNAs are mounted on a common copper plate that has a copper strap routed down the side of each tube to the 4\,K plate. Additional LNAs are installed at 40\,K stage to constitute a two-stage cold LNA system. For more details on the RF component chain, see~\citealt{sath20}.

\subsubsection{Magnetic Shielding}

%

A layer of Cryoperm A4K manufactured by Amuneal\footnote{https://www.amuneal.com/magnetic-shielding/magnetic-shielding-materials} covers the back of the OT to act as a 4\,K magnetic shield, as shown in Figure~\ref{fig:OT_Array_fig}. The A4K magnetic shield extends through the OT to the Lens 2 mounting location shown in Figure~\ref{fig:OT_cutaway}. 
A study has been conducted on the possible UFM magnetic shielding strategies for the microwave SQUID multiplexing readout system~\citep{eve2020}.
As a result of this study, and in order to provide a higher shielding factor for the detector and readout components, we are also investigating plating the OT 1\,K radiation shield with a type-I superconductor.

\begin{figure*}
  \centerline{
    \includegraphics[width=\linewidth]
{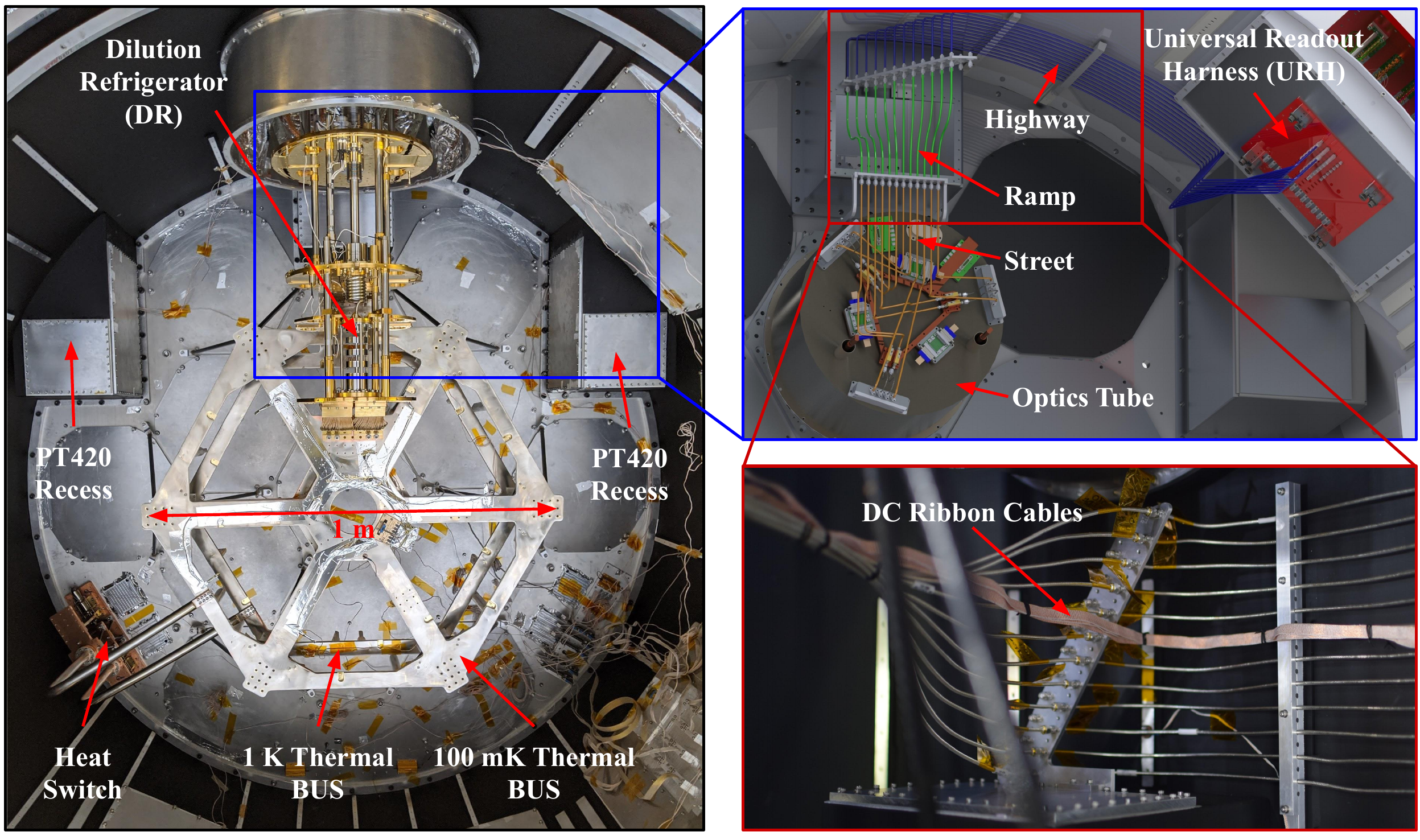}
}
  \caption{4\,K cavity and coaxial cables. The photo on the left shows the overview of the 4\,K cavity, with the 1\,K/100\,mK Thermal BUS, DR, and heat switches installed on the 4\,K plate. The major components are labeled along with 1\,m scale. A zoom-in on the top right shows the design of the coaxial highway and street system for one OT. Other isothermal 4\,K coaxial cable runs and the DR have been hidden for clarity. The highway is in blue, the ramp in green, and the street in orange. Note the relatively short, simple runs of the ramps. The connection to the URH as installed is not as angular as in this rendering. The bottom right photo shows another zoom-in of the installed 4\,K isothermal coaxial cables in the LATR. \label{fig:4K_cavity}}
\end{figure*}

\subsection{\label{subsec:readout_interface_design}Readout Interface Design}

The LATR needs to support $>62,000$ detectors, $>120$ thermometers, 12 heaters, and two heat switches, which all require cables running from the room temperature to cryogenic temperatures. The design requires them to be routed across different temperature stages in an efficient and organized manner. There are five ports on the side of the LATR, penetrating through the 300\,K, 40\,K, and 4\,K stages. Four of the ports are used for detector readout cables while the fifth is used for housekeeping cables (thermometers, heaters, and heat switches).

\vspace{6mm} 

\subsubsection{\label{subsubsec:det_readout}Detector Readout Interface}

In the LATR, 13 OTs require 156 coaxial cables and 2,400 DC wires to be routed without introducing unacceptable thermal loading. To address this, the SO developed the modularized universal readout harness (URH)~\citep{xu/etal:2020b} for both the LATR and the small aperture telescope (SAT)~\citep{ali/etal:2020}. One URH contains 48 coaxial cables penetrating three temperature stages from 300\,K to 4\,K. In addition, each URH carries 600 DC cryogenic wires to bias detectors and amplifiers , and convey flux ramps. The LATR requires four URHs to operate and read out all the detectors in 13 OTs.

Given the complexity of the cable harness and the number of cables bridging different temperature stages, uncontrolled thermal loading is a major concern. MLI blankets were tailored with minimal openings for all the cables on 40\,K and 4\,K plates. Since simulations on the thermal performance of the MLI sheets with many penetrations is unreliable, the URH thermal simulation only includes the thermal conductivity from the cables. The simulated thermal loading is  3\,W at the 40\,K stage and 0.3\,W at the 4\,K stage. Measurement in another cryostat shows loading at the 40\,K stage is ${\sim}$\,7\,W while the loading at the 4\,K stage is ${\sim}$\,0.15\,W. This result demonstrates the success of the MLI design at the 40\,K stage, implying only ${\sim}$\,4\,W of radiative pass through. Meanwhile, the measured 4\,K stage thermal loading is only half of the value predicted by simulations.

All the coaxial cables and the DC wires need to leave the cryostat through the four URHs. Care was taken in the routing of the coaxial cables to satisfy space constraints while minimizing the obstruction of the 4\,K working space by the coax. Additionally, we wanted to reduce the amount of coax that would need to be removed/installed for an OT removal/installation, as well as reduce the complexity of those segments to minimize the amount of time spent removing and installing coax. Therefore, instead of directly routing individual cables for the shortest run, we ran the majority of the cables along the inside of the 4\,K shell, minimizing the interference with the OTs. Analogous to the road system found in the United States, our isothermal 4\,K coax are designed in three parts: highways, ramps, and streets (Figure~\ref{fig:4K_cavity}). The highways run along the inside of the 4\,K shell from the URH to a set of permanently installed bulkheads, each located along the interior of the shell as close as possible to its corresponding OT. The highways are supported along their length, and constitute most of the length of the 4\,K isothermal run (up to $1.5$\,m). Each OT has an individual set of matching bulkheads on the back of the 4\,K shell. The streets are permanently installed sections that run from those bulkheads to penetrations in the magnetic shielding, connecting to other cables deeper in the OT. This arrangement allows most of the complex routing to be permanently installed. Connecting the street bulkheads and the highway bulkheads are the ramp sections: short (${\sim}\,20$\,cm) pieces of hand formed coax. These are the only part of the isothermal 4\,K run that are not permanently installed, and hence are the only pieces that have to be added or removed when installing or removing an OT. The DC cables run along the isothermal 4\,K coax, and are tied down to the coaxial highways, ramps, and streets. See Figure~\ref{fig:4K_cavity} for an example of one OT's isothermal 4\,K run.

\subsubsection{\label{subsubsec:hk_interface_design}Housekeeping Readout Interface}

The LATR's temperatures are monitored with 126 thermometers, distributed across the 80\,K, 40\,K, 4\,K, 1\,K and, 100\,mK stages. At the 80\,K, 40\,K and 4\,K stages, we installed DT-670 silicon diodes, manufactured by Lake Shore Cryotronics.\footnote{\url{https://www.lakeshore.com/}} For the 1\,K and 100\,mK stages, we use  ruthenium oxide (ROX) sensors\footnote{Model \# RX-102A-AA}, also from Lake Shore. The diodes and ROXs were potted in custom-built copper bobbins with Stycast 2850 FT, along with heat sinking wires and a micro connector. All thermometers are read out via a four-wire measurement, with cables manufactured by Tekdata Interconnections LTD.\footnote{\url{https://www.tekdata-interconnect.com/}} The cables are routed via cryogenic breakout boards that were developed for the SO. The thermometers purchased from Lake Shore were calibrated in-house in a dedicated Bluefors DR against pre-calibrated reference sensors.

To carry the signal to the exterior of the LATR, we designed an assembly based on the URH design, called the housekeeping harness. 
Outside of the LATR, we employ Lake Shore measurement modules to do thermometry data acquisition. For 100\,mK thermometers, we utilize two Lake Shore 372 AC Resistance Bridges, each coupled to a 16-channel scanner\footnote{Model 3726}. All other thermometers on higher temperature stages are read out with Lake Shore 240 Series Input Modules. All the thermometry data is read out and stored using the Observatory Control System (OCS) software developed for the SO~\citep{Brian2020}. 
The housekeeping harness is also used to monitor and control other systems including the heat switches, low-power heaters for testing, and high-power heaters to facilitate warming up the system.

\subsection{\label{subsec:tele_interface_design}Telescope Interface Design}

\begin{figure}
    \centerline{
    \includegraphics[width=1.1\linewidth]
{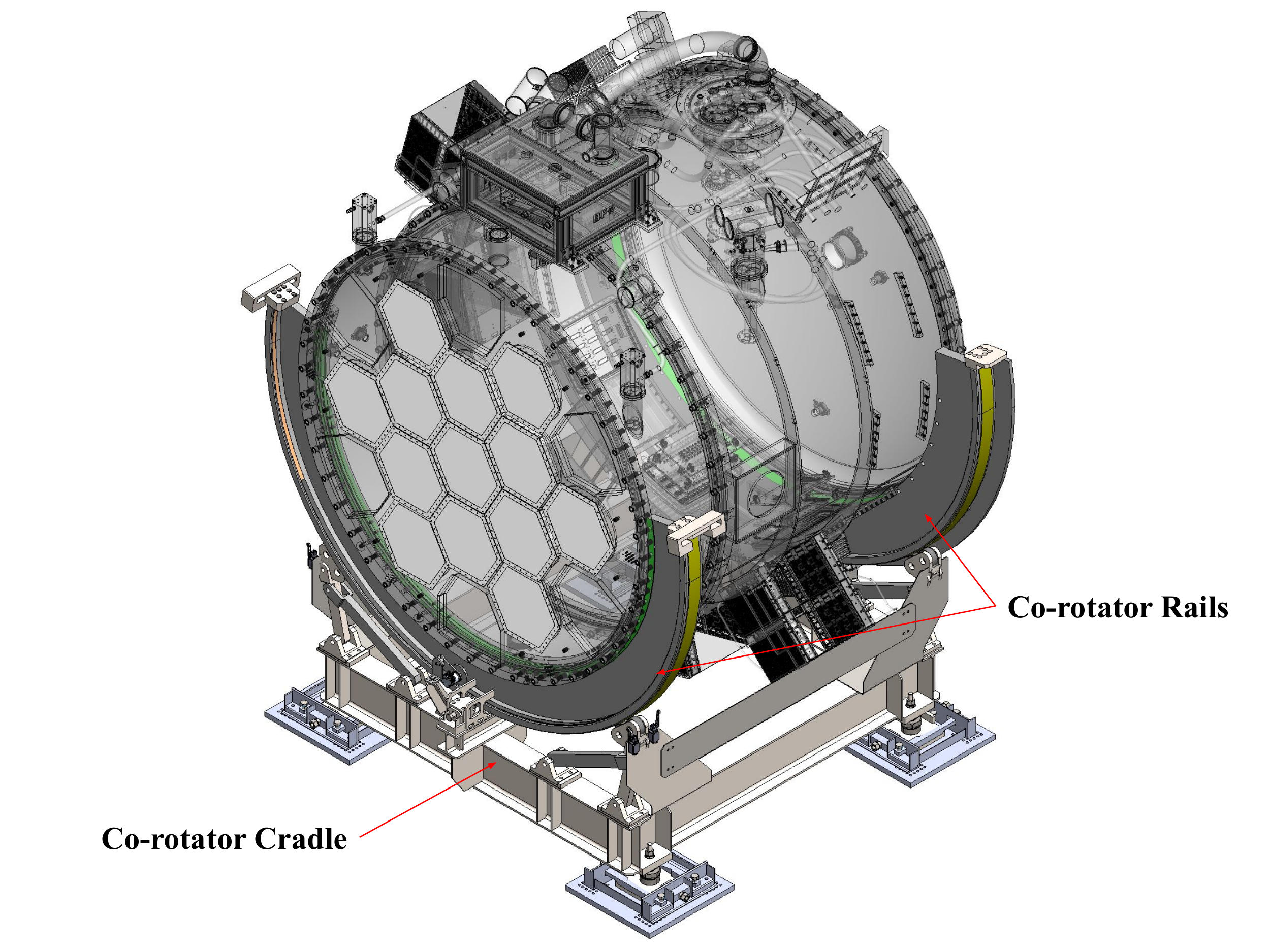}
}
  \caption{The LATR co-rotator. The co-rotator rails are bolted to the front and back flange of the cryostat. The four feet of the co-rotator cradle allows for fine adjustment when aligning the cryostat with the secondary mirror. In this rendering, the cryostat  is transparent to showcase the co-rotator parts.
    \label{fig:corotator}}
\end{figure}

According to the LAT optics design~\citep{dick18}, the LATR must co-rotate with the telescope elevation structure as the telescope points from $0^\circ$ to $90^{\circ}$ in elevation (Figure~\ref{fig:LAT_Xsection}). The co-rotator, as shown in Figure~\ref{fig:corotator}, consists of a pair of co-rotator rails that are bolted to the front and back flange of the LATR, and a cradle that supports and rotates the rails. The design creates a stable support structure for the LATR while allowing adjustments in the co-rotator cradle bases for accurate on-axis alignment. The co-rotation makes sure that the LATR illuminates the same area of the telescope mirrors at different elevations.

The pulse tube cooler hoses, along with all the electronic cables, converge at the cable hub on top of the cryostat (Figure~\ref{fig:latr_external}), before going through the cable wrap.  The cable wrap attaches to the cryostat, and -- in a controlled manner -- bends as the cryostat co-rotates, safely guiding the motion of the hoses and wires  (Figure~\ref{fig:LAT_Xsection}).


\section{\label{sec:latr_validation}LATR Validation and Tests}

During the assembly of the LATR, we performed several rounds of metrology to guarantee the stringent alignment requirements were achieved under different loads. We also measured the resonant frequency of the 1\,K and 100\,mK thermal BUS stages with accelerometers.

The cryogenic properties of the LATR were systematically tested in multiple configurations, including dark tests, optical tests, and tests with OTs installed. We tilted the entire cryostat to test its performance at different orientations as well as to determine the thermal time constant of the system.

After the cryostat and first OT were successfully integrated and tested for the mechanical and cryogenic performance, the RF chain for reading out the detectors was connected (and looped back at the 100\,mK stage). The RF performance will be covered in a future publication.

\subsection{\label{subsec:mechnical_validation}Mechanical Validation}

Comprehensive simulations were performed to set the required tolerance for the alignment of the LATR optical components \citep{dick18}. The location of the LATR is referenced to the telescope-receiver interface: the co-rotator. Utilizing a reference plane that is external to the receiver shell ensures that the optical components within the LATR will be aligned with the primary and secondary mirrors in the LAT. The positions of optical components within the LATR---including the lenses, filters, and detectors---must adhere to tight tolerance requirements, both statically and dynamically (i.e. during co-rotation).

Optimization of Strehl ratios using optical simulations \citep{dick18} lead to tight physical constraints being placed on the locations of our OTs and cryogenic lenses. 
The position of the OTs must be maintained within $\pm3$ mm along the LATR's optical axis and $\pm3$ mm perpendicular to the LATR's optical axis. In addition, the OTs must not exceed $\pm 0.8 ^{\circ}$ in tilt with respect to each other. These constraints need to be met when the cryostat is initially deployed with a small number of OTs and also when it is fully loaded with 13 OTs. The requirement must also be maintained when the LATR is rotated.

To ensure these tight tolerances were met, we utilized the FARO Vantage Laser Tracker\footnote{FARO website: \url{https://www.faro.com/}} to measure the locations of all relevant surfaces both internal to, and external to, the cryostat. The FARO Vantage Laser Tracker has an accuracy of 16 $\mu$m and a single point repeatability of 8 $\mu$m at 1.6 m. 

The Vantage Laser Tracker was used to verify the overall dimensions of the cryostat, the 3D locations of each of the four temperature stages, and the precise locations each OT will inhabit when installed. Due to the nature of the cryostat and the support provided by the 4\,K stage, it was important for us to measure the location of the 4\,K stage both with and without the load of 13 OTs. Upon measuring the various temperature stages with no load on the 4\,K stage, we were able to conclude that the positions of our OTs would be well within our required tolerances. A detailed list of the measured position of the 4\,K plate without OTs can be found in Table~\ref{tab:LATR_4K_metrology}.

\begin{deluxetable}{c c c c}

\tablehead{\colhead{Axis} & \colhead{Tolerance} & \colhead{Deviation from} & \colhead{Deviation from Origin}\\[-3mm]
\colhead{} & \colhead{} & \colhead{ Origin (no load)} & \colhead{(13 OT load)}}

\tablecolumns{4}

\tablecaption{LATR 4\,K Plate Metrology Results\label{tab:LATR_4K_metrology}}

\startdata
    X-axis & 3\,mm & 0.49 $\pm$ 0.03\,mm & 0.31 $\pm$ 0.02\,mm\\
    Y-axis & 3\,mm & 2.04 $\pm$ 0.03\,mm & 2.24 $\pm$ 0.02\,mm\\
    Z-axis & 3\,mm & 2.03 $\pm$ 0.03\,mm & 2.43 $\pm$ 0.02\,mm\\
\enddata
\tablecomments{This table shows the position of the 4\,K plate under two conditions: bare and loaded. Bare measurements refer to those performed with no OT mass dummies installed. Loaded measurements refer to those performed with 13 OT mass dummies installed. Under both conditions, the plate is within the required tolerances provided by \cite{dick18}. The nominal origin, and thus center of the 4\,K plate, is assumed to be (0,0,0) in a 3D coordinate system. Looking at the cryostat from a viewpoint in front of it, Z-axis is the optical axis, Y-axis is pointing towards the ground, and X-axis is pointing towards the right. is  All deviation values are absolute values of the measured deviations of the center of the 4\,K plate from the origin.}
\vspace{-0.5cm}
\end{deluxetable}

We were also able to perform measurements of the 4\,K stage with 13 OT mass dummies installed. These mass dummies accurately reproduced both the total mass and the location of the center of mass for each OT. Once all 13 mass dummies were installed, the metrology process was repeated. Even with all 13 mass dummies added, the 4\,K plate remained within the required positional tolerances. A detailed list of the measurements recorded for the loaded 4\,K plate can be found in Table~\ref{tab:LATR_4K_metrology}. As can be seen from the measurements, the difference between the loaded and the unloaded measurements is less than 0.5\,mm, which is within specifications.  The overall offset in the Y and Z axes is likely due to build-up of manufacturing tolerances.  The Z-axis offset can be removed by re-focusing.  The Y-axis offset could be removed by shifting the 4~K cold plate, if necessary.

\begin{deluxetable}{c c c}

\tablehead{\colhead{Axis} & \colhead{Deviation from Origin} & \colhead{Deviation from Origin}\\[-3mm]
\colhead{} & \colhead{(Simulated)} & \colhead{(Measured)}}

\tablecolumns{3}

\tablecaption{LATR Front Plate Metrology Results - Under Vacuum\label{tab:LATR_FPVacuum_metrology}}

\startdata
    X-axis & 0\,mm & 0.22 $\pm$ 0.05\,mm \\
    Y-axis & 0\,mm & 0.19 $\pm$ 0.05\,mm\\
    Z-axis & 18.1\,mm & 17.2 $\pm$ 0.03\,mm\\
\enddata
\tablecomments{This table highlights the expected and measured position of the LATR's front plate under vacuum. The expected position of the front plate under vacuum is provided by the simulation performed in \citep{orlo18} and shown in Table~\ref{fig:vacuum_fos}. The nominal origin, and thus center of the front plate, is assumed to be (0,0,0) in a 3D coordinate system. All deviation values are absolute values of the measured deviations of the center of the front plate from the origin.}
\vspace{-0.5cm}
\end{deluxetable}

The Vantage Laser Tracker system was also used to measure the LATR's front plate while under vacuum to ensure that the deformation of the plate was within the expected regime. Simulations of the deformation of the front plate can be found in Table~\ref{fig:vacuum_fos} along with the measurement results. From these results, we can see that the measured deformation along the Z-axis, or bowing, was within 1 \,mm of the simulated result (slightly smaller than expected).

The optical simulations performed also provided constraints on our lens positions within the OT themselves. All three lenses within the OTs must maintain a position of $\pm 2$ mm along the LATR's optical axis and $\pm 2$ mm perpendicular to the LATR's optical axis. The tilt of each lens must not exceed $\pm 0.4 ^{\circ}$ with respect to the axis perpendicular to the optical path. The focal plane array within the OT must maintain a position of $\pm 2.5$ mm along the LATR's optical axis. Further constraints placed on the beam guiding 80\,K alumina wedges and the surface accuracy of the lenses can be found in \cite{dick18}.

Due to the size and intricate design of the OTs, we utilized the FARO Edge ScanArm to obtain precise cold optical component and detector array locations. The lens and OT metrology was performed on (warm) tubes  before they were installed in the LATR to allow for easier access to hard-to-reach locations within the tubes. The FARO Edge ScanArm has an accuracy of 34 $\mu$m and a repeatability of 24 $\mu$m at 1.8 m. Each individual mechanical component of every OT is measured and recorded. Partial and full sub-assemblies of the OT are also measured, thus allowing us to understand the position of, parallelism between, distance between, and coaxiality of all components within each OT. Tight constraints on the locations of the focal plane array and lenses created a need for extensive documentation of all dimensions of OT components. Measurements of the positions of the optical elements and their deviations from designed locations can be found in  Table~\ref{tab:LATR_OT_metrology}. From these measurements, we have deduced that all of the lenses, the Lyot stop, and the focal plane in the first optical tube are located within the required tolerances.

\begin{deluxetable}{c c c}

\tablehead{\colhead{Optical} & \colhead{Z-axis} & \colhead{Deviation}\\[-3mm]
\colhead{Element} & \colhead{Tolerance} & \colhead{from Design}}

\tablecolumns{3}

\tablecaption{An example of LATR Optics Tube Metrology Results\label{tab:LATR_OT_metrology}}

\startdata
Lens 1 & 2.0\,mm & 0.47 $\pm$ 0.02\,mm\\
Lens 2 & 2.0\,mm & 0.19 $\pm$ 0.02\,mm\\
Lens 3 & 2.0\,mm & 0.50 $\pm$ 0.02\,mm\\
Lyot Stop & 2.0\,mm & 0.22 $\pm$ 0.02\,mm\\
Detector Arrays & 2.5\,mm & 0.53 $\pm$ 0.02\,mm\\
\enddata
\tablecomments{This table reports the positions of the optical elements within a LATR OT. The z-axis travels along the central axis of the cylindrical OT. The listed tolerances are provided by \cite{dick18}. All deviation values are absolute values of the measured deviations of central plane of each element, perpendicular to the central axis, with respect to the designed central plane for each element.}
\vspace{-0.5cm}
\end{deluxetable}

Since the co-rotator on the LAT is not yet available, we are unable to measure the LATR at a variety of different clocking orientations.   However, the difference between the unloaded and loaded displacements provided in Table \ref{tab:LATR_4K_metrology} suggest that optical alignment will be well within specification.   

Beyond the work on the LATR, the FARO Vantage Laser Tracker will also be used within the LAT to ensure the proper positioning of the mirror panels for the telescope's primary and secondary mirrors. 

\subsection{\label{subsec:cryo_validation}Cryogenic Validation}

In order to know the total thermal loading on each stage precisely, we calibrated all four pulse tube coolers individually, measuring their specific cooling power at different temperatures. The information enabled us to accurately deduce the thermal loading on each stage by closely monitoring the pulse tube cold head temperatures.

The cryogenic validation of the LATR proceeded in incremental steps, starting with the simplest configuration and gradually adding components. We first conducted 4\,K dark tests, with two PT420s and two PT90s. Following completion of this test, we installed the Bluefors DR with the associated thermal straps that coupled the DR to the cryostat 4\,K shell and the 1\,K/100\,mK thermal BUS. From the base temperatures in the dark tests, we backed out the loading on each stage using the cryogenic cooler load curves we calibrated during the preparation. These two rounds of dark tests provide a baseline loading without external radiation.

After the dark tests, we installed 300\,K windows made out of 1/8$''$ ultra high molecular weight polyethylene sheets. Behind each window, a double-sided infrared blocker (DSIR) at 300\,K was installed on the back of the front plate to reflect out-of-band radiative power. Another DSIR filter was installed on the 80\,K stage, followed by an AR coated alumina filter~\citep{nado18} also at 80\,K. The alumina filter absorbs the residual infrared radiation and conducts the heat away efficiently. 
After the alumina filter, another DSIR, at 40\,K, reflects the residual infrared radiation before it enters the 40\,K cavity. The 40\,K cavity then houses the front part of the 13 OTs, including the filters and lenses at 4\,K and below (Figure~\ref{fig:latr_cut}). Prior to installing OTs, metal blanks with MLI were installed on the 4\,K plate to block the radiation from the 40\,K cavity. 

We cryogenically tested the LATR with two windows installed (`2-window' configuration) and subsequently with three windows installed (`3-window' configuration). The `3-window' configuration also had one OT and one URH installed. Given the overall progress of the project, initially we only fabricated three sets of the filters to give us enough fidelity to extrapolate the performance with all 13 windows installed. 

\begin{deluxetable}{c c c c}

\tablehead{\colhead{Test} & \colhead{Filter Plate} & \colhead{Measured} & \colhead{Predicted}\\[-3mm]
\colhead{Configuration} & \colhead{Temperature} & \colhead{Loading} & \colhead{Loading}}

\tablecolumns{4}

\tablecaption{80\,K Stage Thermal Performance\label{tab:80k_loading}}

\startdata
    Dark & 37\,--\,39\,K & 22\,W & 10.1\,W\\
    2-window & 40\,--\,43\,K & 35\,W & 17.9\,W\\
    3-window & 44\,--\,47\,K & 42\,W & 21.8\,W\\
    7-window* & $51$\,K* & 71\,W* & 37.3\,W\\
    13-window* & $65$\,K* & 113\,W* & 60.6\,W\\
\enddata
\tablecomments{Base temperature and thermal loading for the 80\,K stage under different configurations. The temperature range is measured from six thermometers evenly distributed on the 80\,K filter plate. The starred values are extrapolated from measured results. The predicted load is derived from our thermal model in Table~\ref{tab:latr_loading}. Cooling power on the 80\,K stage was designed with a significant margin. The difference between the predicted loading and the measured loading is explained in the text.}
\vspace{-0.5cm}
\end{deluxetable}

Between the 80\,K, 40\,K, and 4\,K stages, the 80\,K is the most sensitive to radiation loading because of the absorptive 80\,K alumina filter. Measurements of the 80\,K stage thermal loading under different configurations are summarized in Table~\ref{tab:80k_loading}. The baseline loading from the dark test is ${\sim}$\,22\,W with each window adding another ${\sim}$\,7\,W radiation loading on the 80\,K stage. If we project the results to 13 windows, the anticipated loading will be ${\sim}$\,113\,W. Given the load curve of the PT90s and thermal strap conductance, we calculate that the 80\,K stage will stay at around 65\,K, safely below the designed 80\,K requirement. Because the 80\,K filter plate is made of thermally-conductive aluminum 1100 series, the temperature gradient across the 2.1\,m diameter plate measured ${\sim}$3\,K with three windows installed. Receiving roughly three times the loading with 13 OTs, the gradient may increase to ${\sim}$\,9\,K, still less than the designed requirement at ${\sim}$\,10\,K (Figure~\ref{fig:thermal_80K}).  

While the observed loading in Table~\ref{tab:80k_loading} meets the LATR's requirements, it is higher than the models predicted.  In the dark configuration, this is likely due to imperfections in the MLI. Considering the size of the 2.1\,m diameter 80\,K filter plate and the complicated geometry of the G-10 tabs, it is unsurprising to observe extra loading due to imperfect shielding of all the surfaces. In the window tests, we measured ${\sim}$\,7\,W per filter set instead of the predicted ${\sim}$\,3\,W. We are still investigating the mismatch, but it is likely a result of higher-than-expected infrared transmission of the DSIR filters. Another factor that should be considered is that all the measurements were performed when both of the pulse tube coolers were operating at $<$\,40\,K, much lower than their nominal 80\,K operation temperature. This is near the lower limit of the temperature the pulse tube coolers can reach, where only a sparsely-sampled calibration curve is available. This adds $\mathcal{O}$(1\,W)-level uncertainties in the measurements.

\begin{deluxetable}{c c c c}

\tablehead{
\colhead{Test} & \colhead{Plate} & \colhead{\multirow{2}{*}{Power}} & \colhead{Predicted}\\[-3mm]
\colhead{Configuration} & \colhead{Temperature} & \colhead{} & \colhead{Loading}}

\tablecaption{40\,K/4\,K Stage Thermal Performance\label{tab:40k_4k_loading}}

\startdata
    \multicolumn{4}{c}{\textbf{40\,K Filter Plate}}\\
    \hline
    Dark & 44\,--\,47\,K & $33_{-1}^{+1}$\,W & 35.3\,W \\
    2-window & 44\,--\,47\,K & $<33$\,W & 35.3\,W \\
    3-window + 1-OT & 44\,--\,48\,K & $<34$\,W & 38.8\,W\\
    \hline
    \hline
    \multicolumn{4}{c}{\textbf{4\,K Plate}}\\
    \hline
    Dark & 3.5\,--\,5.2\,K & $0.8_{-0.1}^{+0.1}$\,W & 0.42\,W \\
    2-window & 3.6\,--\,5.0\,K & $0.8_{-0.1}^{+0.2}$\,W & 0.47\,W \\
    3-window + 1-OT & 3.8\,--\,5.2\,K & $1.3_{-0.2}^{+0.2}$\,W& 0.75\,W\\
\enddata
\tablecomments{Base temperature and loading for the 40\,K filter plate and the 4\,K plate under different configurations, based on measurements at multiple locations. In the `dark' configuration, the two calibrated PT420s cooled the 40\,K and 4\,K stages; starting from the `2-window' configuration, the PT420 on the DR was thermally connected to the main cryostat in 40\,K/4\,K stages. We estimated contribution from the DR PT420 by comparing its temperatures in different configurations to the ones measured in its stand-alone cryostat. We used the average of the two calibrated PT420s and estimated the power given the temperature change on the 4\,K stage. However, the DR PT420 40\,K temperature decreased after it was installed in the main cryostat (being cooled by the other two PT420s), thus we report the number from the other two PT420s as upper limits. In the `3-window' configuration, one OT and one URH were also installed in addition to three windows.}
\vspace{-0.5cm}
\end{deluxetable}

During the dark test, the 40\,K filter plate stayed at 44\,--\,47\,K (measured at six locations on the plate) with an estimated loading of $33\pm1$\,W; the 4\,K plate stayed at 3.5\,--\,5.2\,K (measured at five locations on the plate) with an estimated loading of $0.8\pm0.1$\,W (Table~\ref{tab:40k_4k_loading}). Estimating the thermal loading, especially on the 4\,K stage, is challenging  at these very low power levels where the calibration of the pulse tubes is less certain and thus results in a ${\sim}$\,1\,W loading uncertainty on the 40\,K stage and ${\sim}$\,0.1\,W on the 4\,K stage. To achieve a more accurate measurement, we installed seven heaters (evenly distributed) on the 4\,K plate to add additional power and monitored the temperatures on the stage. From the data, we calculated the thermal conductance of the thermal straps connecting the pulse tube 4\,K cold heads to the cryostat 4\,K plate and then backed out what the thermal loading would be without additional loading from the heaters. This independent measurement gives results consistent with the those from the PT420 load curves.

After the  DR was installed (starting from the `2-window' configuration in Table~\ref{tab:40k_4k_loading}), estimating thermal loading on the 40\,K and 4\,K stage became more difficult because the PT420 on the DR also contributed to the cooling of the 40\,K and 4\,K stages. We did not calibrate the PT420 in the DR since it is deeply integrated in the DR system. Furthermore, the conductivity of the thermal links between the DR 40\,K/4\,K stages and the main cryostat 40\,K/4\,K stages are hard to quantify. The DR and the cryostat 40\,K stages are connected by aluminum coated mylar tape that is intended only for sealing the gaps between them to stop the light leak. The DR and the cryostat 4\,K stages are connected by 5\,$\times$\,10-stacked high-purity (99.999\%) aluminum tabs. The aluminum tabs bridge the ${\sim}$\,6\,mm gap, and have dimensions of $0.5\times50\times75$\,mm. We measured the 40\,K and 4\,K stage temperatures of the DR PT420 and compared to the values previously measured in its designated cryostat as a no-extra-load reference. The DR PT420 4\,K temperature did not change in the `2-window' configuration and rose by ${\sim}$\,0.2\,K in the `3-window' configuration, mainly because of the addition of the OT and the URH. Using the average of the calibrated load curves from the two PT420s, we estimated the additional cooling power due to the DR PT420, and thus the total loading from the observed temperature change. The thermal loading measurements are reported in Table~\ref{tab:40k_4k_loading}. 

Interestingly, the DR PT420 40\,K temperature stage decreased after installation in the LATR. This means that the main cryostat 40\,K stage cools to a lower temperature than the DR 40\,K stage so that it is `lending' cooling power to the DR 40\,K stage when thermally connected. Therefore, we report the 40\,K thermal loading from the two calibrated PT420s as upper limits in Table~\ref{tab:40k_4k_loading}. 

The temperature gradient across the 2.06-m diameter 40\,K filter plate is $\le$\,4\,K with three windows. From the `dark' to the `3-window' configuration, we did not measure significant changes on the 40\,K filter plate in terms of thermal loading and thermal gradient. Based on that observation, the simulation is validated -- showing the 40\,K filter plate will not change significantly with a fully-equipped LATR, largely because only reflective filters are installed on this stage. Additionally, the 40\,K stage of the URH  measured ${\sim}$\,50\,K, 4\,K higher than the 40\,K filter plate average. It is the warmest part of the 40\,K stage since it is both far away from the pulse tube cold heads (Figure~\ref{fig:4K_cavity}) and the source of significant thermal loading (${\sim}\,7$\,W for 4 URHs for the fully loaded configuration). The measured temperatures and the small thermal gradient proved that the pulse tube cooling power is efficiently distributed around the 2-m-long 40\,K stage to maintain the entire stage below 50\,K.

The thermal gradient across the 4\,K plate is around 1.7\,K in the `dark' configuration, and reduced to 1.4\,K after the addition of the DR. The difference between the measured and predicted loading may come from the decreased fidelity of the simulation at the low-power limit. The addition of the windows did not significantly change the loading on the 4\,K which is consistent with the simulation. In the `3-window' configuration, the temperature on the 4\,K shell is ${\sim}$\,5\,K, including the 4\,K part of the URH. This measurement validates that the entire 4\,K stage is efficiently cooled with 3 windows, one OT, and one URH installed, and we see no evidence that the full complement of OTs will introduce unacceptable loading.

Enclosed within the 4\,K cavity, the 100\,mK thermal BUS cooled to $<50$\,mK with the 1\,K thermal BUS maintained at ${\sim}$\,1\,K. Since we do not have 13 OTs during these tests, heaters were installed on both the 100\,mK and 1\,K thermal BUS to simulate thermal loads. With these two heaters, we were able to map out the load curve on the two stages which later informed our understanding of loading from the installed OTs. We also applied the anticipated loading from 13 OTs on the two stages, and the 100\,mK thermal BUS stayed below 100\,mK with $<$10\,mK thermal gradient across its diameter.

The cooldown time for each temperature stage changes with the configuration of the LATR. Intuitively, the more thermal mass we add and the more thermal loading we introduce, the longer the cooldown process will take. Currently, the most comprehensive test we have performed is the configuration with three windows (and the corresponding filters at 300\,K, 80\,K, and 40\,K) and one OT.

\begin{figure}[h] 
    \centering
    \includegraphics[width=0.5\textwidth]{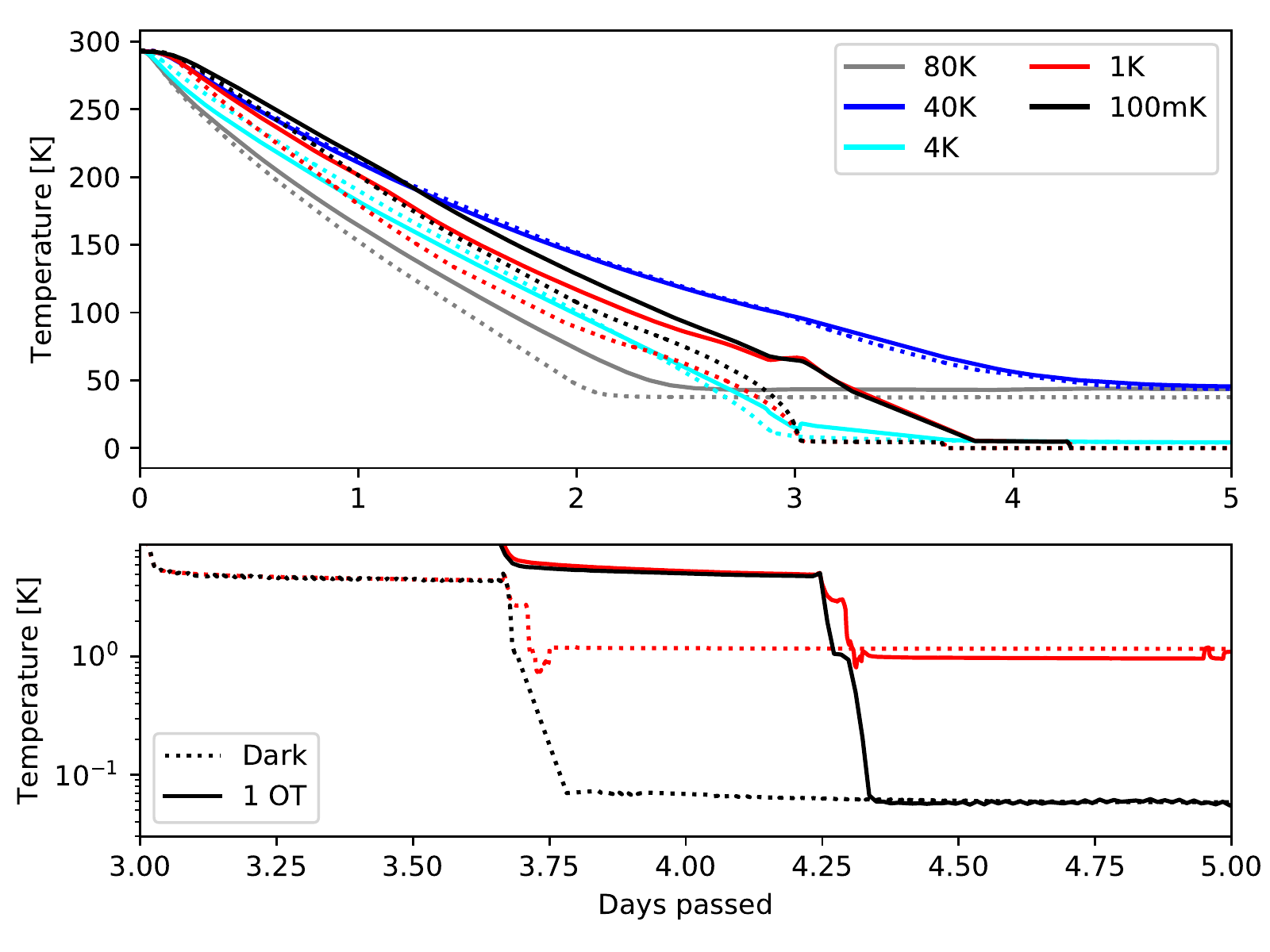}
    \caption{Cooldown curves at each temperature stage from a test run including one OT and three optical windows (and filters).  Although multiple thermometers are installed on each temperature stage, we show only one representative thermometer per stage. Temperatures measured during the dark cooldown are represented by dotted lines, and during the three window cooldown as solid lines. In both configurations, all the stages cooled to base temperatures within 5 days. The bottom panel is a zoomed-in logarithmic plot of the 1\,K and 100\,mK stages. These two stages reached base temperatures within hours of the dilution refrigerator being turned on. }
    \label{fig:Cooldown_curve}
\end{figure}

The first OT was installed into the cryostat behind one of the three window+filter assemblies. During the cooldown, temperatures of different stages as a function of time were recorded. Figure~\ref{fig:Cooldown_curve} shows the cooldown curve from representative thermometers on all temperature stages.


\begin{deluxetable}{c c c c}

\tablehead{\colhead{Temperature}  & \colhead{Cooldown} & \colhead{Plate/BUS} & \colhead{Thermal}\\[-3mm]
\colhead{Stage} & \colhead{Time} & \colhead{Temperature} & \colhead{Gradient}}

\tablecaption{LATR Cooldown Time\label{tab:cooldown_temp}}
\startdata
    80\,K & 3\,days & 43\,K & $\pm 1$\,K \\
    40\,K & 5\,days & 47\,K  & $\pm 3$\,K \\
    4\,K & 4\,days & 4.5\,K & $\pm 0.7$\,K\\
    1\,K & 5\,days & 1.2\,K & $\pm 0.07$\,K \\
    100\,mK & 5\,days & 55\,mK & $\pm 10$\,mK\\
\enddata
\tablecomments{Cooldown time measured in the three windows configuration with one OT installed. This is not to be confused with the simulated 13 OTs cooldown, which will take significantly longer due to added cold mass.}
\vspace{-5mm}
\end{deluxetable}


In this configuration, the 80\,K stage was able to reach its base temperature within 3 days and the 4\,K stage was able to reach its base temperature within 4 days, as shown in Table~\ref{tab:cooldown_temp}. After the 4\,K stage reached its base temperature, the DR began operating, cooling the 1\,K and 100\,mK stages to base temperatures within several hours (Figure~\ref{fig:Cooldown_curve}). The 40\,K stage, specifically the 40\,K filter plate, reached base temperature in 5 days because it is the farthest from the PT420 cold heads (Figure~\ref{fig:latr_cut}). The observed cooldown times of the various stages with a 4K cold mass were expected and reflected the simulated performance. Also shown in Table~\ref{tab:cooldown_temp} is the base temperature for each stage---measured either on the plate or on the BUS---along with the corresponding thermal gradient.

Thermal loading on the 100\,mK stage was increased by $<$4\,$\mu$W after installing one OT in the center position. With the amount of added power so small compared to the rest of the 100\,mK structure, it is hard to obtain a more accurate loading value from a single OT. With the addition of active readout components and optical loading during operation, the total expected loading from one OT is $\lesssim$5\,$\mu$W~\citep{xu/etal:2020b, Katie2020}. Extrapolating from this number, the estimated extra loading from 13 OTs will be ${\sim}$\,70\,$\mu$W. Since we did not have all 13 OTs available, we installed a heater on the 100\,mK thermal BUS stage and dissipated 65\,$\mu$W to simulate anticipated thermal loading. Although concentrating 13 OTs loading to one location will not reflect the actual loading distribution, the purpose of this test was to ensure the proper performance of the 100\,mK thermal BUS. The 100\,mK thermal BUS was designed to effectively distribute the cooling power to 13 OTs. When the 70\,$\mu$W total loading was applied, the thermal BUS temperature rose to 78\,mK at the coldest point (closest to the cold strap) and 87\,mK at the hottest point (furthest from the cold strap). In order to ensure that these temperatures were acceptable, we also needed to inspect the thermal gradient between the thermal BUS and OT 100\,mK stage. The target temperature for a focal plane baseplate in the fully loaded 13 OT configuration is $<$100\,mK. We observed that the gradient between thermal BUS and OT 100\,mK stage was ${\sim}$\,5\,mK. This means that when we extrapolate the loading to 13 OTs, the warmest OT focal plane stage will be at most 92\,mK, accounting for the uncertainty in our measurement of the 100\,mK loading from one OT. This implies that the LATR will meet the required thermal specifications.
 
\begin{figure}[h] 
    \centering
    \includegraphics[width=0.5\textwidth]{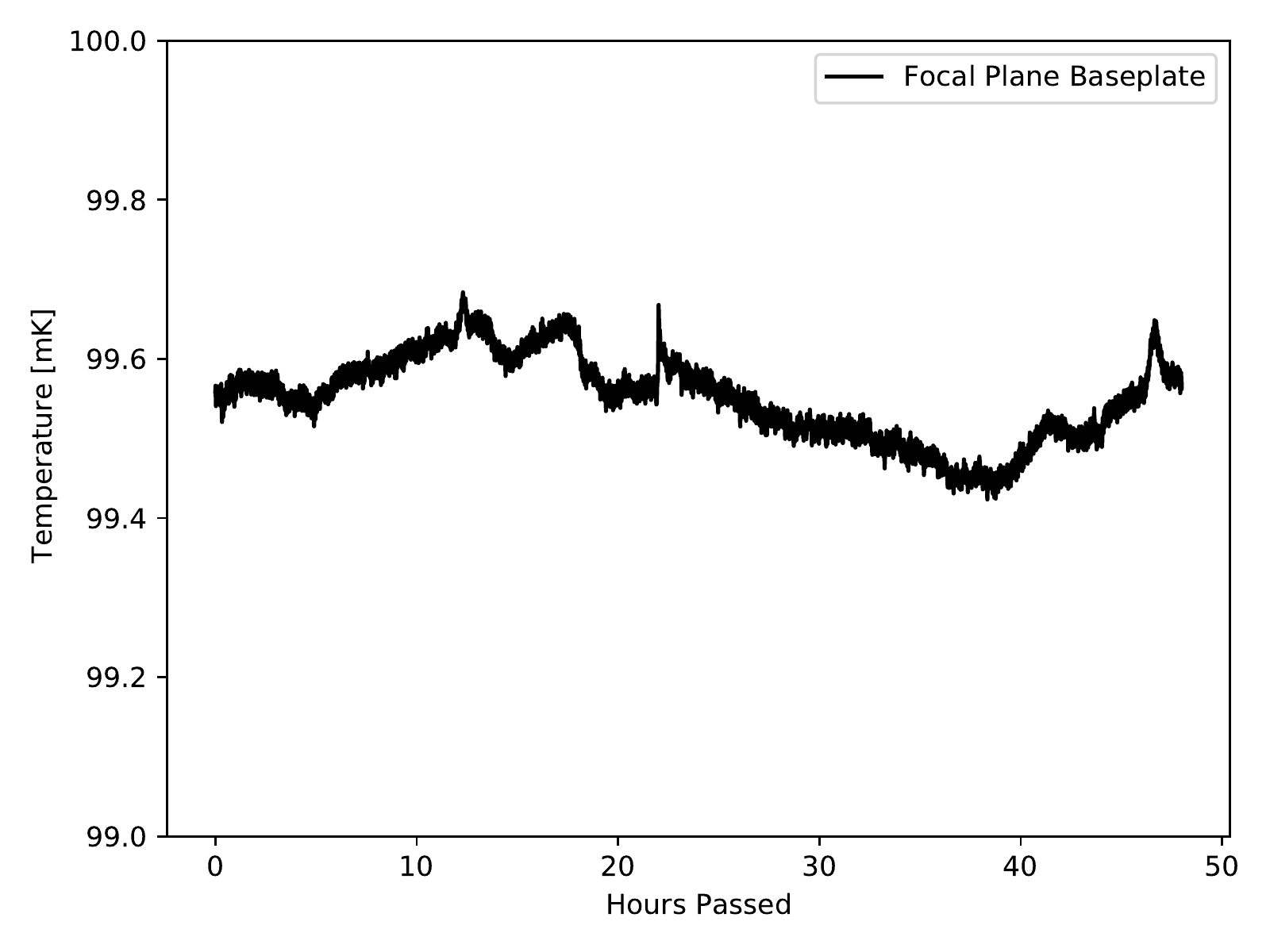}
    \caption{The temperature of an OT 100\,mK stage over 48 hours. On this timescale, the OT focal plane stage temperature varies by $\lesssim$0.3\,mK. At this level, bath temperature fluctuations will not meaningfully detract from overall detector stability.}
    \label{fig:FPB_timestream}
\end{figure}

 After ensuring that the LATR would be able to thermally support 13 OTs, we moved onto testing the thermal performance of the OTs themselves. In order to maintain the required thermal environment for the detector arrays, we needed to ensure that OT 100\,mK stage had both long-term temperature stability and a negligible thermal gradient across the stage. Long-term temperature stability is important to detector performance, as bath temperature drifts can impact detector data quality. Consistent thermal bath temperature between adjacent detector arrays is also important to ensure the best quality of detector data. In Figure \ref{fig:FPB_timestream}, we show how the temperature of one OT 100\,mK stage changes over the course of 48 hours. Similar to the thermal tests described thus far, the OT was dark -- with the 4\,K and 1K stages blanked-off. Constant power was applied to a heater near the DR to raise the temperature of the OT 100\,mK stage to the fiducial operating temperature of ${\sim}$\,100\,mK. As can be seen in Figure \ref{fig:FPB_timestream}, the cold plate temperature varied by $\lesssim$0.25\,mK over the (representative) 48 hour period. At this level, bath temperature fluctuations will have negligible impact on detector stability. The thermal gradient across the OT 100\,mK stage was measured to be $\lesssim$1\,mK at base temperature. A gradient this small means that adjacent detector arrays within a single OT will have negligible differences in bath temperature, preventing degradation of the detector data quality. Thus, these validation test show that we have met the design requirements. 

\subsection{\label{subsubsec:vibration_tests}Vibration Tests}

An area of concern in the thermal-mechanical design of the receiver is microphonic heating of the coldest stages, and particularly of the detector arrays, by vibrational sources in the cryostat and the receiver cabin \citep{Bhatia1996}. In addition, vibration pick up can add to detector noise. Predicting the magnitude of this loading prior to the construction of the receiver is very difficult, and moreover the heating is most noticeable at the 1\,K and 100\,mK stages. FEA can provide some leverage in the design phase by allowing us to predict which components of the 1\,K and 100\,mK stages will have low frequency fundamental vibrations modes that are more likely to couple to the vibrational environment of the receiver. These simulations were performed, and are detailed in \cite{orlo18}. Ultimately, however, vibrational FEA of the entire receiver is not feasible,  and the FEA contains no information about the amplitude of the vibrations. Therefore, we need to test the transfer function of the receiver in the lab.

To test the vibrational sensitivity of the 100\,mK stages, we attached a Buttkicker mini Concert transducer\footnote{https://thebuttkicker.com/buttkicker-mini-concert/} to the front flange of the cryostat, driven using its associated amplifier with ${\sim}$\,200\,W of power. We tested attaching the Buttkicker to the middle and back flanges and found it made no significant difference. We used a tone generator to supply the signal to the amplifier. We then monitored the temperature at a number of points on the  100\,mK stages while slowly sweeping the input tone. By carefully recording when we started the sweep, the speed of the sweep, and the times at which we observed spikes in the temperatures, we could determine the frequency at the time of the spike, which we interpreted as a resonance frequency somewhere in the system. We ran the sweeps both increasing and decreasing in frequency as a consistency check and to account for any time delays in the temperature response. Once we identified a resonance, we would sweep very slowly over the resonance to accurately recover its frequency. Generally all these tests agreed to within about 1\,Hz. Further, by observing the magnitude of the temperature rise at a number of locations, we could roughly localize the source of the heating and infer which part of the cryostat was strongly coupled to that vibration. By this method, we have determined that the lowest harmonics of the 100~mK stage are at 21, 24, 27, and 31\,Hz, all of which are associated with the structure at or near the OT 100mK cold finger. The cold finger shown in Figure~\ref{fig:OT_cutaway} is a long and thin rigid body, so it is particularly susceptible to microphonic heating.  The resonance at 24~Hz is by far the largest in amplitude, at approximately 7 times larger temperature gain than the others. While current calculations indicate this heating mode will not affect the performance of the receiver, avenues for stiffening the part are being explored.


\section{\label{sec:future_development}Future Development}

Although the tests described in this paper demonstrate that the LATR meets all requirements, there are some areas where the system performance could be further optimized. Some of them were over-designed, meaning we could have achieved the same performance without sacrificing performance on other aspects; on the other hand, some of the parts were under-designed, meaning the performance could be boosted significantly if we had given more margins in the design. We hope that the lessons learned here will be beneficial for future experiments developing large-volume ultra-cold cryostats, for example CMB-S4~\citep{s4tb17, s4sb16}.


The 40\,K stage, running from the very back to the very front of the cryostat, is the most spatially extended stage in the LATR. The shells in this stage are responsible for conducting the cooling power from the pulse tube cold heads, which are mounted at the midpoint of the stage. The initial design called for the 40\,K shell to be made of 1/8$''$ aluminum 1100 series, which proved to be not thick enough. It caused a gradient concentrated around the location where the pulse tubes are attached to the 40\,K shell. The thermal conductivity  was improved by installing additional 1/4$''$-1/2$''$ thick high-purity (99.999\%) aluminum cladding on the 40\,K shells. The modification significantly reduced the thermal gradient on the 40\,K stage. In hindsight, it would have been wise to leave extra margins for thermal conductivity, particularly on a stage that spatially spans the entire cryostat.

More generally, a lesson we learned from developing and testing the LATR is that distributing the cooling power is more important than adding cooling power. For future large cryostat development efforts, evenly distributing the cooling power by boosting the thermal conductivity between different parts, should be a high priority in the cryogenic design. With well-thermally-connected temperature stages, not only will thermal gradients be small, but the cooldown time will be significantly reduced, removing the need for other complicated pre-cooling systems.

A significant amount of effort went into minimizing the number of bolted joints on 1\,K and 100\,mK stages, in order to reduce the temperature gradient on these two stages. We accomplished this by welding vertical rods onto the 1\,K and 100\,mK thermal BUSs to reach as close to the OT 1\,K and 100\,mK cold fingers as possible. While the 1\,K BUS performed well enough to meet the specification, the 100\,mK BUS suffered from unreliable welding quality, which resulted in a large temperature gradient between the OT focal plane base plate and the cryostat's 100\,mK BUS. To remedy this, we chose to drill new holes and installed bolted rods on the 100\,mK BUS, which performed much better than the welded rods. Due to the size of the 100\,mK BUS, a large amount of heat is required for the welding process, which can lead to warping and difficulty controlling the quality of the welded joints. For future applications, we would recommend employing bolted joints for ease of manufacture and installation. 

\begin{figure}
    \centering
    \includegraphics[width=0.7\columnwidth]{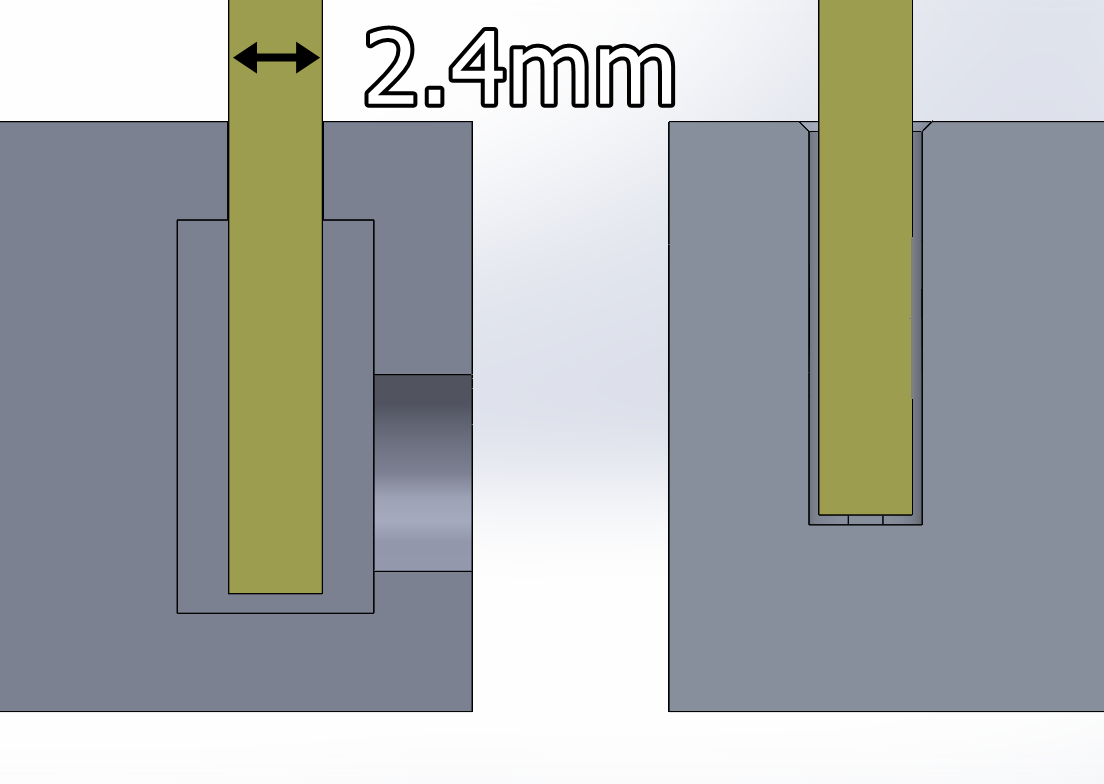}
    \caption{Recommended (left) and installed (right) G-10 tab glue joint designs, shown as a cross-section of the tab foot. The scale is slightly different between the two but the G-10 tab (yellow) is the same size. The aluminum foot is in grey. The cavity to the left in the recommended design is a glue injection passage. Note the overhang by the aluminum foot of the glue join in the recommended design, and the much tighter fit of the G-10 tab.}
    \label{fig:glue_change}
\end{figure}

Given the results of the pull test of the glue joint, for future development we would recommend changing the design of the glue joints. In our current design, our G-10 tabs are inserted into a U-shaped cavity with pre-filled glue and then left to cure. This design has the advantage of simplicity in manufacture and assembly, but has the significant disadvantage of relying on the bond strength of the glue. Indeed, in all pull testing we performed the glue failure was adhesive, specifically from the aluminum walls of the cavity. Instead, we recommend a 'pocket-cavity' design, wherein a pocket is machined in the foot, with a tight clearance hole for the G10 at the pocket entrance. Injection holes around the sides provide access to the cavity for gluing. This design is more complicated to manufacture but has the advantage that the glue can no longer fail in adhesion from the aluminum due to the 'roof' of the cavity. An example of this design is shown in Fig.~\ref{fig:glue_change}. While we did not switch the design of the G-10 tabs, as they were already installed when we developed this design, we did successfully use a similar design for the 1\,K to 100\,mK standoffs.

\section{\label{sec:conclusion}Conclusion}

The Simons Observatory (SO) Large Aperture Telescope Receiver (LATR) will be installed in the Large Aperture Telescope (LAT) that utilizes a cross-Dragone design with two 6\,m aperture mirrors. The LATR contains five cryogenic stages, including 80\,K, 40\,K, 4\,K, 1\,K, and the ultimate 100\,mK stage where detectors operate. Various cryogenic optics, including optical filters and cryogenic lenses, are installed on the cryogenic stages to achieve designed thermal and optical performances. At its heart, the LATR is capable of cooling $>62,000$ transition edge sensors (TES)~\citep{irwin/hilton:2005} to below 100\,mK, with a sub-mK long term temperature stability. At this temperature, detector noise is sub-dominant to incoming photon shot noise. 

The detectors and the $\le$\,4\,K optics are packaged in modular units called optics tubes (OTs). The modularized design facilitates efficient installation and removal without sacrificing precise control of the relative positions of the optics.
Our initial plan is to deploy with seven OTs in the nominal LATR configuration with the potential upgrade to 13 OTs in the future. 
Carrying detector data out of the cryostat requires hundreds of coaxial cables and thousands of cryogenic wires. To organize all the cables and make it easily serviceable, the SO developed the universal readout harness to modularize the 300\,K to 40\,K wiring.

The LATR has been constructed and extensively tested. We started with the simplest configuration and gradually added more factors. Currently, the LATR has been tested with three optical windows, one OT, and two feedthroughs installed (one for detector signals and one for housekeeping). All of the five cryogenic stages cool to their base temperatures within five days (the cooldown time will become longer as the number of OT increases), with the anticipated thermal loading and thermal gradient on each temperature stage. Extrapolating from these partially-equipped results, the LATR will perform as designed when all the 13 OTs and 5 feedthroughs are installed.

The development of the LATR offers critical insight and experience on sub-Kelvin cryostats at this unprecedented size, which will be valuable for future CMB experiments, including CMB-S4~\citep{s4sb16, s4tb17}.

\begin{acknowledgements}

We thank the Vertex group who have been working closely with us on the telescope and receiver interface design. Dynavac, Meyer Tool\footnote{https://www.meyertool.com/}, PVEng, and Fermilab cryogenic engineering group have provided a number of helpful suggestions regarding the cryostat design and manufacture process. We thank Jeffrey Hancock and Harold Borders from University of Pennsylvania Physics machine shop. 


This work was supported in part by a grant from the Simons Foundation (Award \#457687, B.K.).
We acknowledge support from the JSPS KAKENHI Grant Number JP17H06134.

\end{acknowledgements}

\bibliography{main}{}
\bibliographystyle{aasjournal}



\end{document}